\documentclass[draft]{agujournal2019}
\usepackage{url} 
\usepackage{lineno}
\usepackage{trackchanges} 
\usepackage{soul}
\usepackage{amsmath}
\usepackage{bm}

\soulregister\ref7
\soulregister\cite7
\soulregister\citeA7

\let\div\relax
\DeclareMathOperator\grad{\bm{\nabla}}
\DeclareMathOperator\div{\bm{\nabla}\cdot}

\DeclareMathOperator\Pe{\text{Pe}}
\DeclareMathOperator\St{\text{St}}

\newcommand{\pd}[2]{\dfrac{\partial #1}{\partial #2}}

\draftfalse

\journalname{JGR: Planets}

\begin{document}

%
%


\title{Compositional layering in Io driven by magmatic segregation and volcanism}

%
%




\authors{Dan C. Spencer\affil{1}, Richard F. Katz\affil{1}, Ian J. Hewitt\affil{2}, David A. May\affil{1}, and Laszlo P. Keszthelyi\affil{3}}


\affiliation{1}{Department of Earth Sciences, University of Oxford, Parks Road, Oxford OX1 3PR, UK}
\affiliation{2}{Mathematical Institute, University of Oxford, Woodstock Road, Oxford OX2 6GG, UK}
\affiliation{3}{U.S. Geological Survey, Astrogeology Science Center, 2255N. Gemini Dr., Flagstaff, AZ 86001, USA}




\correspondingauthor{Dan Spencer}{dan.spencer@earth.ox.ac.uk}




\begin{keypoints}
\item We present a model of Io that couples crust and mantle dynamics to a simplified compositional system.
\item Magmatic segregation and volcanism cause rapid stratification, leading to the formation of refractory melts in the lower mantle.
\item Io's highest temperature eruptions can be explained as deep refractory melts that migrate to the surface.
\end{keypoints}

%
%

%
%


\begin{abstract}
The compositional evolution of volcanic bodies like Io is not well understood. Magmatic segregation and volcanic eruptions transport tidal heat from Io's interior to its surface. Several observed eruptions appear to be extremely high temperature ($\geq 1600~$K), suggesting either very high degrees of melting, refractory source regions, or intensive viscous heating on ascent. To address this ambiguity, we develop a model that couples crust and mantle dynamics to a simple compositional system. We analyse the model to investigate chemical structure and evolution. We demonstrate that magmatic segregation and volcanic eruptions lead to stratification of the mantle, the extent of which depends on how easily high temperature melts from the more refractory lower mantle can migrate upwards. We propose that Io's highest temperature eruptions originate from this lower mantle region, and that such eruptions act to limit the degree of compositional stratification.
\end{abstract}

\section*{Plain Language Summary}
Io is vigorously heated by the tides it experiences from Jupiter. This heating causes the interior to melt, feeding volcanic eruptions onto the surface. When a rock is heated, some chemical components enter the melt at lower temperatures than others. In this work we use a new model to show that low-melting-point magmas form and rise toward the surface, leaving behind a deep mantle composed of high-melting-point rock. This deep high-melting-point rock eventually melts and must also rise upward in order to allow the lower mantle to lose heat. We propose that high-temperature magmas formed in the deep mantle can rise all the way to the surface, providing an explanation for the highest temperature eruptions.

%
%

%


%
%
%
%

\section{Introduction}
Jupiter's moon Io is the most volcanically active body in the solar system. Its volcanism is a result of tidal heating from its mean motion resonance with Europa and Ganymede, which causes widespread melting in its interior \cite{peale_melting_1979,oreilly_magma_1981}. The export of this tidal heat through the crust by a volcanic system is a process commonly referred to as `heat-piping' \cite{oreilly_magma_1981}. Despite its long history of study, it is not well known to what extent melting and volcanism control Io's interior structure and evolution and, in particular, if these processes create compositional layering within the mantle. Constraints on interior structure would be provided by measurements of the composition and temperature of erupted lavas. To keep pace with recent improvements in observational techniques (e.g. \citeA{davies_determination_2016,davies_novel_2017,de_kleer_ios_2019,de_kleer_variability_2019}), interior evolution models that are predictive of eruption temperatures and compositions are increasingly required.

\citeA{keszthelyi_magmatic_1997} presented an initial attempt to estimate the geochemical and petrological structure of Io's interior that would arise from the extensive volcanism. They predicted that the crust would be dominated by felsic lavas rich in incompatible elements and that the mantle would be dominantly a forsterite-rich dunite. When the initial Galileo observations suggested widespread eruption of ultramafic lavas and constrained the temperature of the Pillan eruption to $1870\pm 25~$K \cite{mcewen_high-temperature_1998}, this model was abandoned. It was replaced by a model that called upon a region with $\sim 50\%$ partial melting at the base of the crust. This configuration hypothetically allowed efficient recycling of the erupted lavas back into the mantle \cite{keszthelyi_revisiting_1999,keszthelyi_post-galileo_2004}. This magma-ocean model was supported by Galileo magnetometer results \cite{khurana_evidence_2011} and is consistent with the suggestion of magnesian orthopyroxenes in Ionian lavas \cite{geissler_global_1999}. The magma-ocean model predicts a well-mixed and geochemically homogeneous mantle \cite{keszthelyi_post-galileo_2004}; erupted lavas would be largely uniform in temperature and composition, most likely similar to terrestrial komatiites \cite{williams_komatiite_2000}.

However, there were significant challenges to the magma-ocean model as proposed in \citeA{keszthelyi_post-galileo_2004}. For example, once partial melting exceeds $\sim 20$\%, the shear modulus drops to the point that tidal dissipation cannot match the surface heat flow \cite{moore_tidal_2003,bierson_test_2016,renaud_increased_2018}, limiting the possible thickness of such a high-melt-fraction layer (however, dissipation in a magma ocean may be significant \cite{tyler_tidal_2015,hay_powering_2020}). Furthermore, applying a different thermal model to the Pillan eruption, its temperature was revised down to $\sim 1600~$K \cite{keszthelyi_new_2007}. Indeed, even the initial \citeA{mcewen_high-temperature_1998} results showed most eruptions being consistent with $\sim 1300~$K (i.e., basaltic) temperatures. Spectroscopic constraints on the mineralogy of Io's lavas were always known to be weak because the Galileo camera did not observe far enough into the infrared to reliably detect other key minerals such as olivine \cite{geissler_global_1999}. These issues led to a revised magma-ocean model with the maximum degree of mantle partial melting only reaching $\sim25\%$ and decreasing rapidly with depth \cite{keszthelyi_new_2007}. Auroral hotspot oscillations have been used as evidence against a magma ocean \cite{roth_constraints_2017}, and reanalysis of the magnetometer results suggests that plasma interactions with the atmosphere provide an alternative explanation to a magma ocean \cite{blocker_mhd_2018,de_kleer_tidal_2019}. More recently, \citeA{spencer_magmatic_2020} showed that high melt fractions can arise within a decompacting boundary layer at the top of a low-melt-fraction mantle. Indeed, the distinction between a magma-ocean model and a low-melt-fraction model has significantly reduced since \citeA{keszthelyi_magmatic_1997} and \citeA{mcewen_high-temperature_1998}; at this point, the hypothesis that Io is a largely solid body that has developed significant stratification needs to be investigated. 

In this work we present a fluid dynamical model of crust and mantle dynamics that builds on the recent work of \citeA{spencer_magmatic_2020} by including compositional evolution. The compositional model is in the form of a two-component phase diagram between hypothetical refractory and fusible components. We use this simplified theory to investigate the effect of magmatic segregation and volcanic eruptions on leading-order chemical structure. Our results show that magmatic segregation causes a rapid stratification of the mantle, with fusible material in the upper mantle and crust, and refractory material at depth. Magma forms in both the upper and lower mantle and, importantly, magma must be able to leave the lower mantle in order to facilitate heat loss. The model exhibits two distinct modes of behaviour, depending on the fate of magma produced in the lower mantle. If lower mantle melts stall within the upper mantle, high temperature eruptions should not occur. However, if these refractory melts migrate to the surface, they can provide an explanation for the highest temperature eruptions observed on Io.

The manuscript is organised as follows. First we outline the physics of the model before presenting results showing the two distinct modes of behaviour. We demonstrate the time evolution of both modes, and investigate the effect of bulk composition on the system. We then discuss these results in the context of present and potential future observations.

\section{Model description}
\begin{figure}
    \centering
    \includegraphics[scale=0.15]{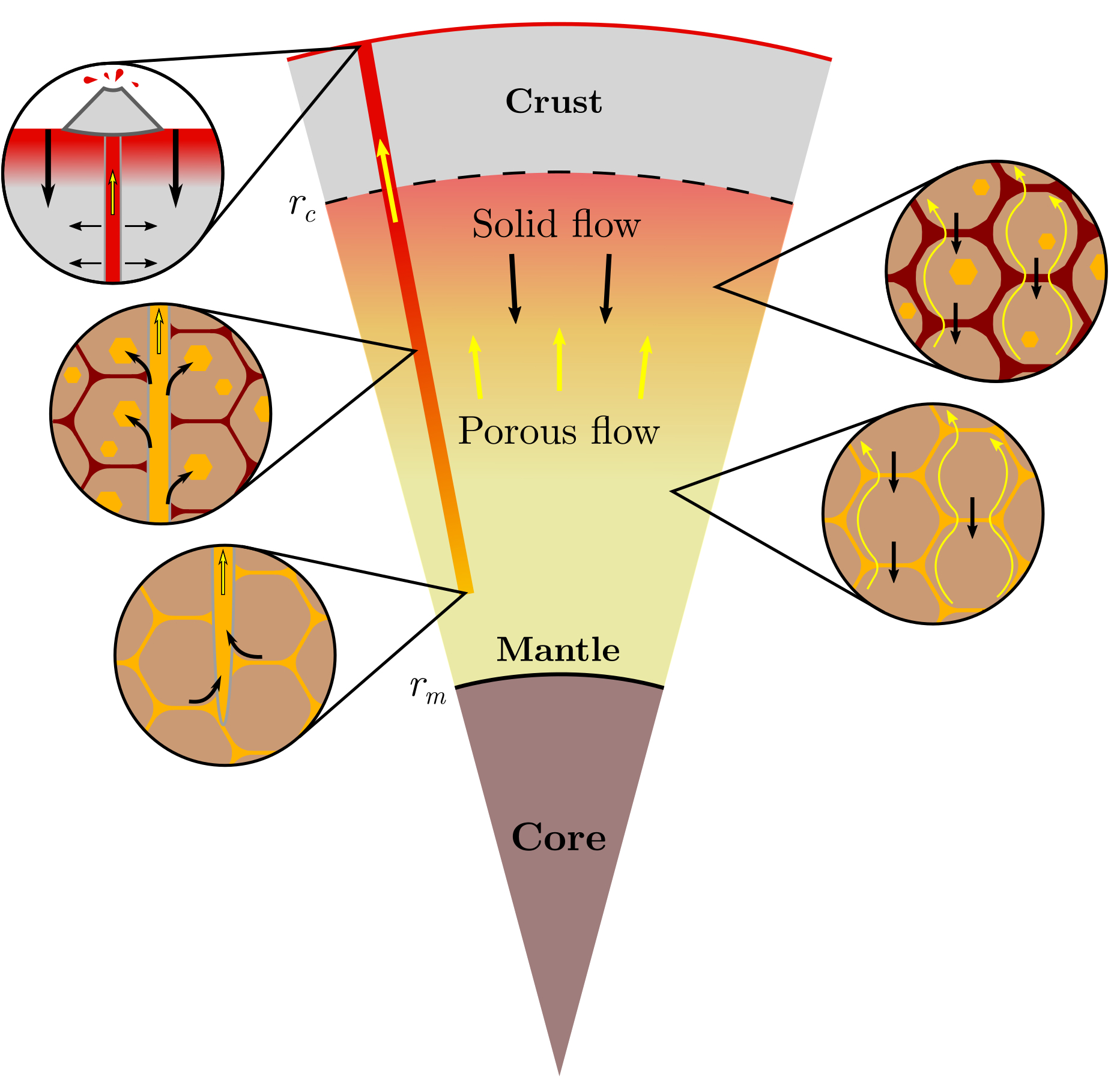}
    \caption{Schematic of the model. Magma rises buoyantly in the mantle while the solid moves downwards. If a critical overpressure is exceeded, magma is extracted to a magmatic plumbing system. It freezes (is emplaced) from the plumbing system back into the continuum at a rate defined in equation \eqref{eq:M}. Some magma reaches the surface, fueling volcanic eruptions and burying the crust. The composition of erupted magma determines the composition of the crust. The core is excluded from the model.}
    \label{fig:schematic}
\end{figure}

The model, shown schematically in figure \ref{fig:schematic}, considers the evolution and dynamics of a tidally heated body composed of a mixture of two chemical components. It is an extension of that described in \citeA{spencer_magmatic_2020} using the same equations, and also solved in one dimension. Here it is extended to consider conservation of chemical species and the effect of composition on melting behaviour, using a phase diagram described below. We consider the crust and mantle to be a continuum that can either be entirely solid or partially molten, depending on the local energy content, and solve a system of conservation equations for mass, momentum, energy, and chemical species.

Alongside the continuum, we model a magmatic plumbing system that provides a means of upward magma transport distinct from magmatic segregation. \citeA{keszthelyi_magmatic_1997} proposed that deep, refractory magmas may sometimes ascend to the surface from great depth, but a mechanism to allow this has not been explored. We assume that anywhere magma reaches high overpressure, it enters into a magmatic plumbing system and migrates upward; this system can be present in both the mantle and the crust. The plumbing system could be interpreted as a system of fractures formed by buoyant, high pressure melt. However, in the formulation of our model we are purposefully agnostic to its exact physical form; we consider possible interpretations in section \ref{section:discussion}. When magma enters the plumbing system, it transports the local melt composition and temperature upward into the upper mantle and crust. The flux of plumbing-system melt that reaches the surface is the erupted flux; its composition sets the composition of the newly resurfaced crust. As in \citeA{spencer_magmatic_2020}, the crust is defined as the portion of the domain that is below the solidus (where the porosity is zero), and so the thickness of the crust is the distance over which cold, surface material downwells before it is heated sufficiently to begin re-melting.

\begin{figure}
    \centering
    \includegraphics[scale=0.23]{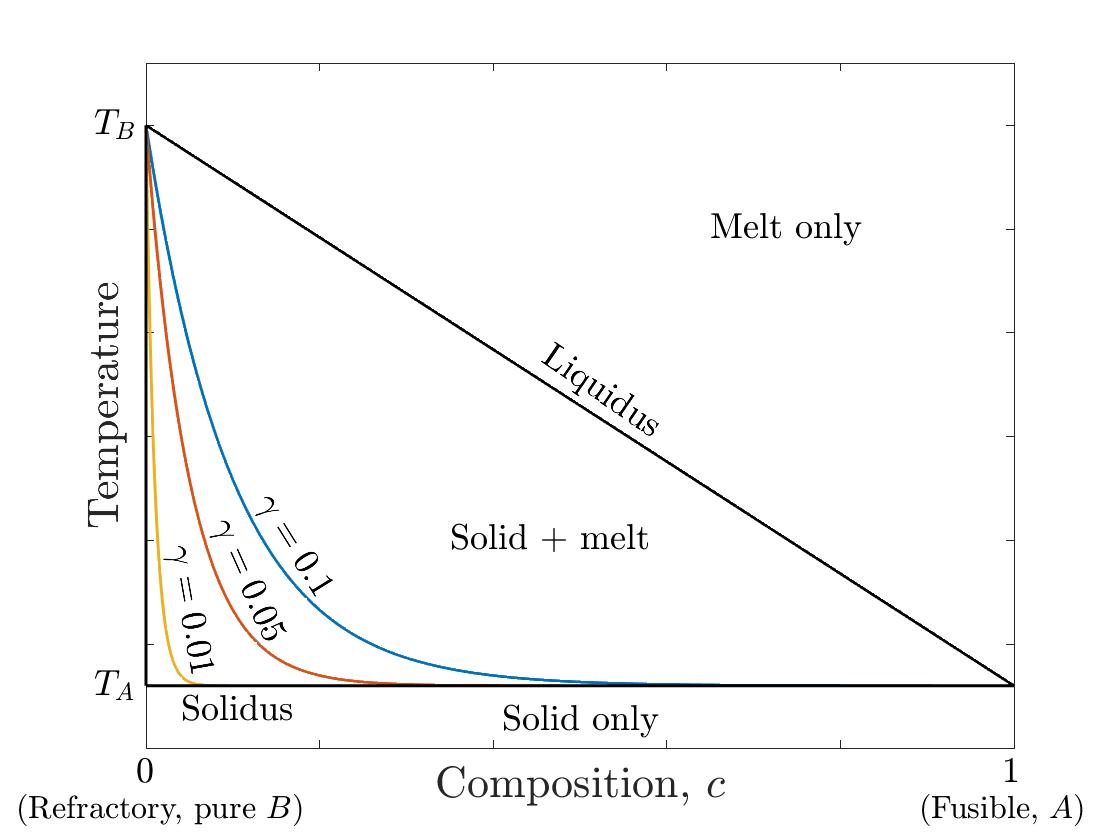}
    \caption{The phase diagram employed in the model. The black lines show the solidus and liquidus between a refractory component $B$ and a fusible component $A$. Coloured lines show the smoothed solidus using equation \eqref{eq:solidus} for different values of $\gamma$, which allow the presence of a small amount of fusible material in solid solution with component $B$. As $\gamma \rightarrow 0$, the smoothed solidus approaches the solidus of pure solid $B$. The full model uses a smoothed solidus with $\gamma = 0.01$, and the reduced model uses the $\gamma = 0$ solidus.}
    \label{fig:eutectic}
\end{figure}

We revisit the thermochemical melting models that have been used to predict the segregation of Io's mantle into an upper fusible layer and a deep layer of almost-pure olivine \cite{keszthelyi_magmatic_1997}. Our approach is to simplify the compositional model to two representative end-members, aiding their incorporation into a dynamical framework. We consider Io to be composed of a mixture of these two components, with a melting behaviour that is described by the two-component phase diagram shown in figure \ref{fig:eutectic}. The presence of fusible material (component $A$) significantly reduces the melting point of the refractory component (component $B$), and so upon heating, fusible melts are produced until component $A$ is almost entirely removed from the system. These types of compositional model have proven fruitful in studies of mantle melting at mid-ocean ridges \cite{katz_porosity-driven_2010,katz_consequences_2012}.

As in \citeA{spencer_magmatic_2020}, we assume spherical symmetry motivated by the global distribution of Io's volcanoes \cite{kirchoff_global_2011,williams_volcanism_2011}. The one-dimensional approach of this work precludes our ability to investigate processes such as thermochemical convection that may be a consequence of stratification, a point we discuss in section \ref{section:conv}, below. We focus our analysis on the chemical evolution of the system, and therefore take tidal dissipation to be uniform, avoiding dependence on poorly constrained rheological parameters \cite{bierson_test_2016,renaud_increased_2018}. In actuality, the tidal heating rate depends on radius, latitude, and longitude, and so while we would not expect significant changes in radial structure to arise from its inclusion \cite{spencer_magmatic_2020}, it is likely to be an important component of models that aim to  predict surface variability. We neglect the pressure-dependence of the melting temperature due to the small size of Io and hence the low pressures in the mantle. We also neglect solid-state phase change and any compositional dependence of latent heat or phase density. For more detailed petrological modelling, it may be important to include these effects.

Our model considers the time-dependent evolution of the interior structure and composition, and explores the evolution to a steady state. We develop a reduced model to elucidate key features of the dynamics predicted by the full model. The reduced model is formulated at steady state and its structure is motivated by solutions obtained to the full model; it is detailed in \ref{appendix:reduced}.


\subsection{Model equations}
We consider a generic refractory component $B$ and a fusible component $A$, and the phase diagram shown in figure \ref{fig:eutectic}. The concentration of the fusible component $A$ in phase $i$ (solid $s$ or liquid $l$) is denoted $c_{i}$, and that of the refractory component is $1 - c_{i}$. The solidus temperature $T_{s}$ is given by
\begin{linenomath*}
\begin{equation}
    T_{s} = T_{B} + (T_{A}-T_{B})\frac{1-e^{-c_{s}/\gamma}}{1-e^{-1/\gamma}},
    \label{eq:solidus}
\end{equation}
\end{linenomath*}
and the liquidus temperature $T_{l}$ is given by
\begin{linenomath*}
\begin{equation}
    T_{l} = T_{B} - (T_{B}-T_{A})c_{l},
    \label{eq:liquidus}
\end{equation}
\end{linenomath*}
where $T_{B}$ is the melting point of the refractory component, $T_{A}$ is the melting point of the fusible component, and $\gamma>0$ is a parameter that controls the amount of fusible material that is incorporated in a solid solution with component $B$. We allow this small degree of solid solution simply because it provides a smoothed solidus curve, which facilitates our numerical method (the effect of smoothing the solidus is small, and is discussed in \ref{appendix:reduced}). As $\gamma \rightarrow 0$, the smoothed solidus approaches that of pure refractory component $B$. The chosen form for the solidus should not be interpreted as representative of any underlying thermodynamics.

The model of \citeA{spencer_magmatic_2020} is described by conservation equations for mass, momentum, and energy in a compacting two-phase medium and conservation of mass and energy equations in the magmatic plumbing system. These are
\begin{linenomath*}
\begin{equation}
    \div(\bm{u}+\bm{q}) = - E + M,
    \label{eq:mass_cont}
\end{equation}
\end{linenomath*}
\begin{linenomath*}
\begin{subequations}
\begin{gather}
    \bm{q} = -\frac{K_{0}\phi^{n}}{\eta_{l}}\left[ (1-\phi)\Delta \rho \bm{g} + \grad P\right], \label{eq:momA} \\
    P = \zeta \left( \div \bm{u} - M\right), \label{eq:momB}
\end{gather}
\end{subequations}
\end{linenomath*}
\begin{linenomath*}
\begin{equation}
    \frac{1}{\rho C}\pd{H}{t} + \div [(\bm{u}+\bm{q})T] + \div \left[(\phi \bm{u} + \bm{q})\frac{L}{C}\right] = \div (\kappa \grad T) + \frac{\psi}{\rho C} - E\left(T+\frac{L}{C}\right) + M\left(T_{p}+\frac{L}{C}\right),
    \label{eq:energy_cont}
\end{equation}
\end{linenomath*}
\begin{linenomath*}
\begin{equation}
    \div \bm{q}_{p} = E-M,
    \label{eq:plum_mass}
\end{equation}
\end{linenomath*}
\begin{linenomath*}
\begin{equation}
    \div (\bm{q}_{p}T_{p})  = ET - MT_{p},
    \label{eq:energy_plum}
\end{equation}
\end{linenomath*}
where $\bm{u}$ is the solid velocity, $\bm{q}=\phi(\bm{v}_{\textrm{liquid}}-\bm{u})$ is the Darcy segregation flux, $E$ is the extraction rate to the plumbing system, and $M$ is the emplacement rate from the plumbing system. Porosity is denoted by $\phi$, and $K_{0}\phi^{n}$ is the permeability, in which $n$ is the permeability exponent. In addition, $\Delta \rho$ is the density difference between solid and liquid, $\bm{g}=-g\hat{\bm{r}}$ is the gravity vector, $\eta_{l}$ is the liquid viscosity, $P=(1-\phi)(P_{\textrm{liquid}}-P_{\textrm{solid}})$ is the compaction pressure, and $\zeta=\eta/\phi$ is the compaction viscosity, related to shear viscosity $\eta$. Bulk enthalpy is defined as $H=\rho CT+\rho L\phi$, $T$ is temperature, $L$ is the latent heat, $C$ is the specific heat capacity, $\rho$ is the density, $\psi$ is the volumetric tidal heating rate, $\kappa$ is the thermal diffusivity, $T_{p}$ is temperature in the plumbing system, and $\bm{q}_{p}$ is the plumbing system flux.

Conservation of mass \eqref{eq:mass_cont} tells us that material leaves the crust--mantle system by extraction to the plumbing system and enters the crust--mantle system by emplacement from the plumbing system back into the continuum. We note that ``emplacement" may have different interpretations in other works, but here it simply means the arrest and freezing of rising plumbing-system melts within the interior. Conservation of momentum is formulated by the combination of Darcy's law \eqref{eq:momA}, which tells us that fluid flow is driven by buoyancy and compaction pressure gradients, with the compaction relation \eqref{eq:momB}, which relates the liquid overpressure to the compaction rate $\div \bm{u}$ \cite{mckenzie_generation_1984}. Equation \eqref{eq:momB} includes magmatic emplacement because we assume that emplacement does not cause fluid pressurisation. Conservation of energy \eqref{eq:energy_cont} tells us that changes in bulk enthalpy occur by the advection of sensible and latent heat, diffusion of sensible heat, tidal heating, the energy removed by extraction, and the energy delivered by emplacement. We note that in \citeA{spencer_magmatic_2020} bulk enthalpy was normalised by the volumetric heat capacity $\rho C$. Conservation of mass \eqref{eq:plum_mass} in the plumbing system tells us that the plumbing system flux increases when material is extracted from the mantle and decreases when material is emplaced back into the continuum. Equation \eqref{eq:energy_plum} represents conservation of energy in the plumbing system. There are no time derivatives in equations \eqref{eq:plum_mass}--\eqref{eq:energy_plum} because the plumbing system is assumed to occupy negligible volume.

To the equations above, we add an equation that tracks the composition of the system
\begin{linenomath*}
\begin{equation}
    \pd{\overline{c}}{t}+\div [(\phi \bm{u}+\bm{q})c_{l}] + \div [(1-\phi)\bm{u}c_{s}] = - Ec_{l} + Mc_{p},
    \label{eq:cnc_bulk}
\end{equation}
\end{linenomath*}
where $\overline{c}=\phi c_{l} + (1-\phi)c_{s}$ is the phase averaged composition and $c_{p}$ is the composition of material in the plumbing system. This equation tells us that changes in phase averaged composition occur through advection of the liquid composition, advection of the solid composition, extraction of the liquid to the plumbing system, and emplacement of the plumbing system material. We neglect compositional diffusion due to the large advective velocities compared to chemical diffusivity. The composition of plumbing system material is given by a conservation of chemical mass equation
\begin{linenomath*}
\begin{equation}
    \div (\bm{q}_{p}c_{p}) = Ec_{l} - Mc_{p},
    \label{eq:mass_plum_A}
\end{equation}
\end{linenomath*}
where the plumbing system composition can only change by the addition of melts from the crust--mantle system of a different composition.

As in \citeA{spencer_magmatic_2020} we assume that the emplacement rate of magma from the plumbing system to the continuum is proportional to the temperature difference between the plumbing system material and the local continuum
\begin{linenomath*}
\begin{equation}
    M = \begin{cases}
    \frac{h_{M}C(T_{p}-T)}{L} \quad & T \geq T_{A}, \\
    \frac{h_{C}C(T_{p}-T)}{L} \quad & T_{A} > T \geq T_{e}, \\
    0 \quad & T<T_{e},
    \end{cases}
    \label{eq:M}
\end{equation}
\end{linenomath*}
where $T_{e}$ is an elastic limit temperature below which no emplacement occurs \cite{spencer_magmatic_2020}. The emplacement rate constant $h$ is discussed at length in \citeA{spencer_magmatic_2020}, but here we propose that it may have different values in the mantle $h_{M}$ and the crust $h_{C}$ (the crust is where $T<T_{A}$). The mechanisms by which magma propagates through a partially-molten medium are likely to be very different to those in a solid, and so would be expected to have a different efficiency of magma transport. In this work, $h_{C}$ is directly analogous to $h$ in \citeA{spencer_magmatic_2020} and the behaviour with different values of $h_{M}$ will be explored.

Extraction of liquid from the mantle into the plumbing system is treated in the same way as in \citeA{spencer_magmatic_2020}; the transfer is taken to be a function of liquid overpressure,
\begin{linenomath*}
\begin{equation}
    E = \begin{cases}
    \nu (P-P_{c}) \quad & P\geq P_{c}, \\
    0 \quad & P<P_{c},
    \end{cases}
    \label{eq:E}
\end{equation}
\end{linenomath*}
where $\nu$ is an extraction rate constant (units s$^{-1}$Pa$^{-1}$), and $P_{c}$ is a critical overpressure that the liquid must exceed in order to be extracted into the plumbing system. We recall that $P$ is the overpressure relative to the lithostatic pressure $P_{\textrm{solid}}$, not the absolute liquid pressure $P_{\textrm{liquid}}$. We take $P_{c}$ to be a constant, but a more realistic model might relate this parameter to depth and the local system state, to capture the different pressures required to initiate and sustain dikes.

The full model to be solved comprises equations \eqref{eq:mass_cont}--\eqref{eq:E}, which govern the time evolution of temperature, porosity, and composition, as well as the magma and solid velocities. The phase averaged composition $\overline{c}$ and the bulk enthalpy $H$ uniquely define the temperature, porosity, and liquid and solid compositions through the solidus and liquidus equations \eqref{eq:solidus}--\eqref{eq:liquidus}, the definition of bulk enthalpy, and the definition of phase averaged composition. The boundary conditions state that there is zero solid and liquid velocity and zero heat flux at the base of the mantle ($r_{m}$ in figure \ref{fig:schematic}), and that there is a prescribed surface temperature $T_{s}$. The composition at the surface is set by the erupted composition, which together with the zero basal fluxes, conserves the bulk composition. The bulk composition is therefore effectively set by the initial conditions.

Parameter values and definitions are given in table \ref{table:parameters1}. The system is scaled (see \ref{appendix:scaling}) and spherical symmetry is assumed so that all variables are a function of only radial position $r$ and time. The system is solved using the Portable, Extensible Toolkit for Scientific computation (PETSc) \cite{petsc-efficient,petsc-web-page,petsc-user-ref,katz_numerical_2007}. Details of the implementation are given in \ref{appendix:implementation}. The code is benchmarked against the single-chemical-component model in \citeA{spencer_magmatic_2020}.

\begin{table}[ht!]
\caption{Dimensional parameters}
\centering
\begin{tabular}{l l l l}
\hline
Quantity & Symbol & Preferred Value & Units \\
\hline
Radial position & $ r $ & & m \\
Radius & $R$ & $1820$ & km \\
Core radius$^{1}$ & $r_{m}$ & $700$ & km \\
Crustal radius & $r_{c}$ & & m \\
Boundary layer coordinate & $ Z $ & & m \\
Solid velocity & $ u $ & & m/s \\
Segregation flux & $ q $ & & m/s \\
Volcanic plumbing flux & $ q_{p} $ & & m/s \\
Porosity & $ \phi $ & & \\
Permeability constant$^{2}$ & $ K = K_{0}\phi^{n}$ & $10^{-7}$&m$^{2}$ \\
Permeability exponent$^{2}$ & $n$ & 3 & \\
Density & $\rho $ & $3000$ & kg/m$^{3}$ \\
Density difference & $\Delta \rho $ & $500$ & kg/m$^{3}$ \\
Gravitational acceleration & $g$ & $1.5$ & m/s$^{2}$ \\
Shear viscosity & $ \eta $ & $ 1\times 10^{20} $ & Pa\,s \\ 
Liquid viscosity & $\eta_{l}$ & $1$& Pa\,s \\
Volume transfer rate & $\Gamma$ & & s$^{-1}$ \\
Emplacement rate$^{3}$ & $M$ & & s$^{-1}$ \\
Crustal emplacement constant$^{*}$ & $h_{C}$ & $5.7$ & Myr$^{-1}$ \\
Mantle emplacement constant & $h_{M}$ & & Myr$^{-1}$ \\
Extraction rate$^{3}$ & $E$ & & s$^{-1}$ \\
Extraction constant$^{3}$ & $\nu$ & $1.4\times 10^{-5}$ & Myr$^{-1}$Pa$^{-1}$ \\
Compaction pressure & $P$ & & MPa  \\
Critical overpressure$^{3}$ & $P_{c}$ & $0$ & MPa \\
Compaction viscosity & $\zeta$ & & Pa\,s \\
Bulk enthalpy & H & & J/m$^{-3}$ \\
Temperature & $T$ & & K \\
Plumbing system temperature & $T_{p}$ & & K \\
Solidus temperature & $T_{s}$ & & K \\
Liquidus temperature & $T_{l}$ & & K \\
Solidus constant & $\gamma$ & $0.01$ & \\
Elastic limit temperature$^{3}$ & $T_{e}$ & $1000$ & K \\
Refractory melting temperature & $T_{B}$ & $1500$ & K \\
Fusible melting temperature & $T_{A}$ & $1230$ & K \\
Surface temperature & $T_{\textrm{surf}}$ & $150$ & K \\
Latent heat & $L$ & $4\times 10^{5}$ & J/kg \\
Specific heat capacity & $C$ & $1200$ & J/kg/K \\
Phase-averaged composition & $\overline{c}$ & & \\
Solid composition & $c_{s}$ & & \\
Liquid composition & $c_{l}$ & & \\
Plumbing system composition & $c_{p}$ & & \\
Tidal heating rate$^{**}$ & $\psi$ & $4.2\times 10^{-6}$ & W/m$^{-3}$ \\
\hline
\multicolumn{4}{l}{ $^{*}$ $h$ in \citeA{spencer_magmatic_2020}} \\
\multicolumn{4}{l}{ $^{**}$ Such that the integrated heating matches the observed input$^{4}$ of $\sim 1\times 10^{14}~$W} \\
\multicolumn{4}{l}{$^{1}$\citeA{bierson_test_2016}, $^{2}$\citeA{katz_magma_2008}, $^{3}$\citeA{spencer_magmatic_2020},} \\
\multicolumn{4}{l}{ $^{4}$\citeA{lainey_strong_2009}} \\
\end{tabular}
\label{table:parameters1}
\end{table}

\section{Results}
The steady-state behaviour of the model across parameter space can be broadly divided into two distinct modes. This division is on the basis of the transport of refractory melts that form in the lower mantle, which is controlled by the value of the mantle emplacement constant $h_{M}$. The results in this section are framed to exhibit the contrasting behaviour of these two modes; the implications of each mode will be discussed further below. In mode 1, rising refractory magma in the magmatic plumbing system interacts and exchanges substantial energy with the lower-temperature partially-molten upper mantle. This drives all plumbing-system magmas to freeze within the upper mantle and, as a result, refractory melts to not reach the crust. In mode 2, refractory plumbing-system magmas rise through the upper mantle with little to no interaction. These melts reach the base of the crust, combine with more fusible melts, and are erupted to the surface. Figures \ref{fig:SS_example} and \ref{fig:mass_comp_schematic} show steady-state solutions for the full model for each of the two modes. Figure \ref{fig:t_evolve} shows the evolution of the model from an initial uniform state, again for each of the two modes. Finally, in figure \ref{fig:cbulk_vary} we summarise the behaviour of the model as a function of the bulk composition of the body, demonstrating the transition between the two modes. These figures are discussed further below.

In this paper we do not explore the parameter space of the crustal emplacement constant $h_{C}$, the elastic limit temperature $T_{e}$, nor the critical extraction pressure $P_{c}$. The effect of variation in these parameters was considered by \citeA{spencer_magmatic_2020} and their effects here are the same. The crustal emplacement constant $h_{C}$ and the elastic limit temperature $T_{e}$ control the thickness and temperature distribution in the crust, and the critical extraction pressure $P_{c}$ affects the melt fraction in decompacting boundary layers that occur where magma is extracted to the plumbing system. In the results presented here, we choose values of $h_{C}$ and $T_{e}$ that give reasonable crustal thicknesses and temperature distributions. We take $P_{c}=0$ and explore whether compositional effects also exert a control on melt fractions.

\subsection{Two Modes of Magmatism}
\begin{figure}[ht!]
    \centering
    \includegraphics[width=\linewidth]{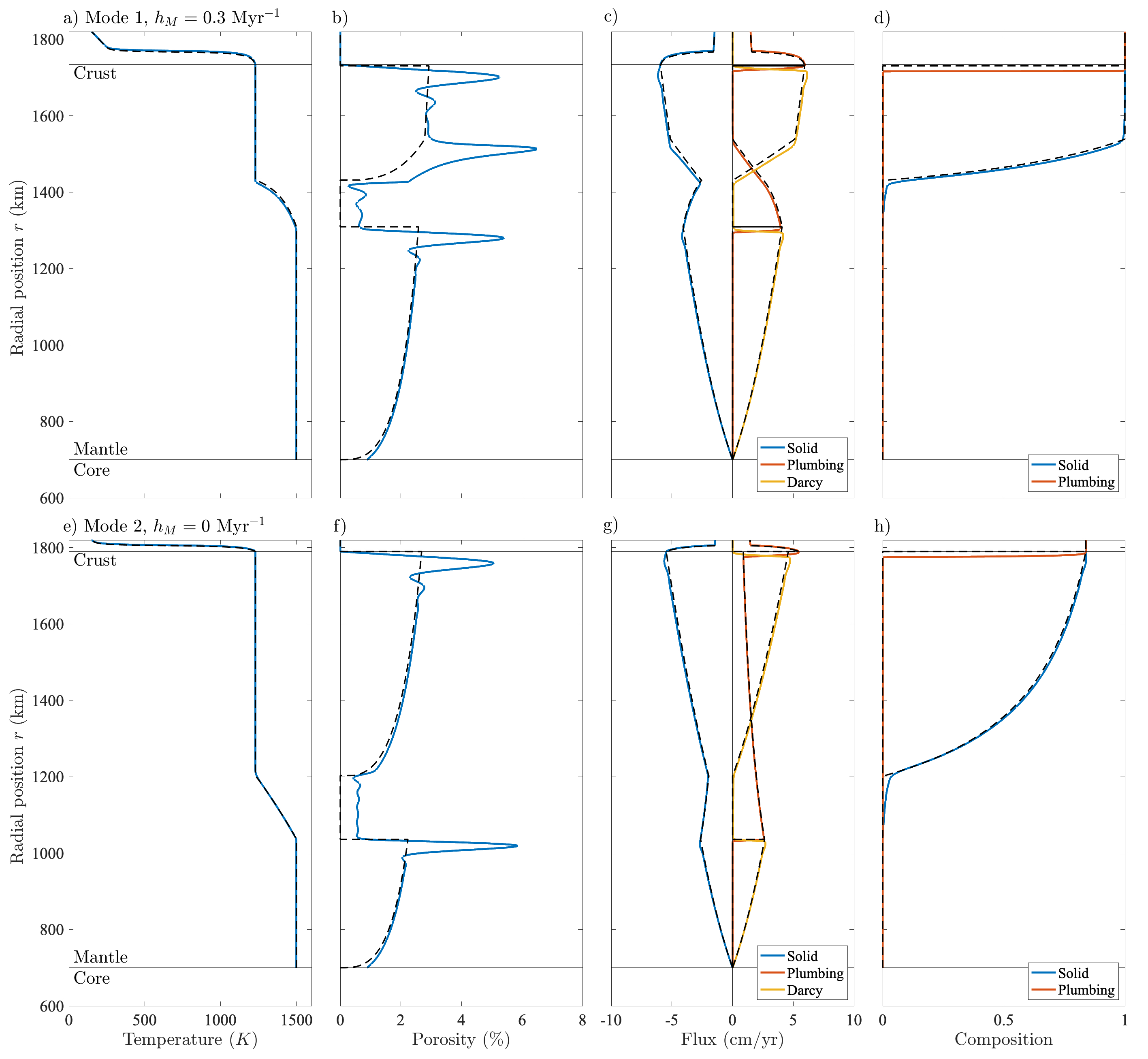}
    \caption{Steady-state solutions to the full model for two end-member behaviours showing temperature; porosity; solid, plumbing, and Darcy fluxes; solid and plumbing system compositions. Panels a--d show mode 1 where $h_{M}=0.3~$Myr$^{-1}$; deep refractory plumbing material is emplaced into the upper mantle. Panels e--h show mode 2 where $h_{M}=0$; deep refractory material is not emplaced in the mantle. Bulk composition is 0.5. In both modes the lower mantle is segregated to a purely refractory composition at temperature $T_{B}$, but in mode 2 the ability of refractory material to migrate to the crust means that the upper mantle is a mixture of refractory and fusible components. In mode 1 the emplacement of refractory melts into the upper mantle drives increased melting, resulting in a porosity peak in the lower part of the upper mantle. The dashed lines show solutions to the reduced model. Parameter values are given in table \ref{table:parameters1}.}
    \label{fig:SS_example}
\end{figure}

Figure \ref{fig:SS_example} shows temperature, porosity, fluxes, and compositions at steady state for two representative values of $h_{M}$. Refractory magmas that form in the lower mantle are transferred to the magmatic plumbing system at the top of the lower mantle, enabling their continued rise. As they rise through the upper mantle, they are emplaced at a rate proportional to $h_{M}$, and it is the size of this parameter that distinguishes the two modes. Mode 1 arises when $h_{M}$ is sufficiently large that all the melt from the lower mantle is emplaced into the mid- and upper mantle. Mode 2 arises when some of the melt extracted from the lower mantle reaches the crust, which occurs if $h_{M}$ is sufficiently small. Solid lines in figure \ref{fig:SS_example} are steady-state solutions to the full model; dashed lines are solutions to the reduced model (see \ref{appendix:reduced}).

\begin{figure}[ht!]
    \centering
    \includegraphics[width=\linewidth]{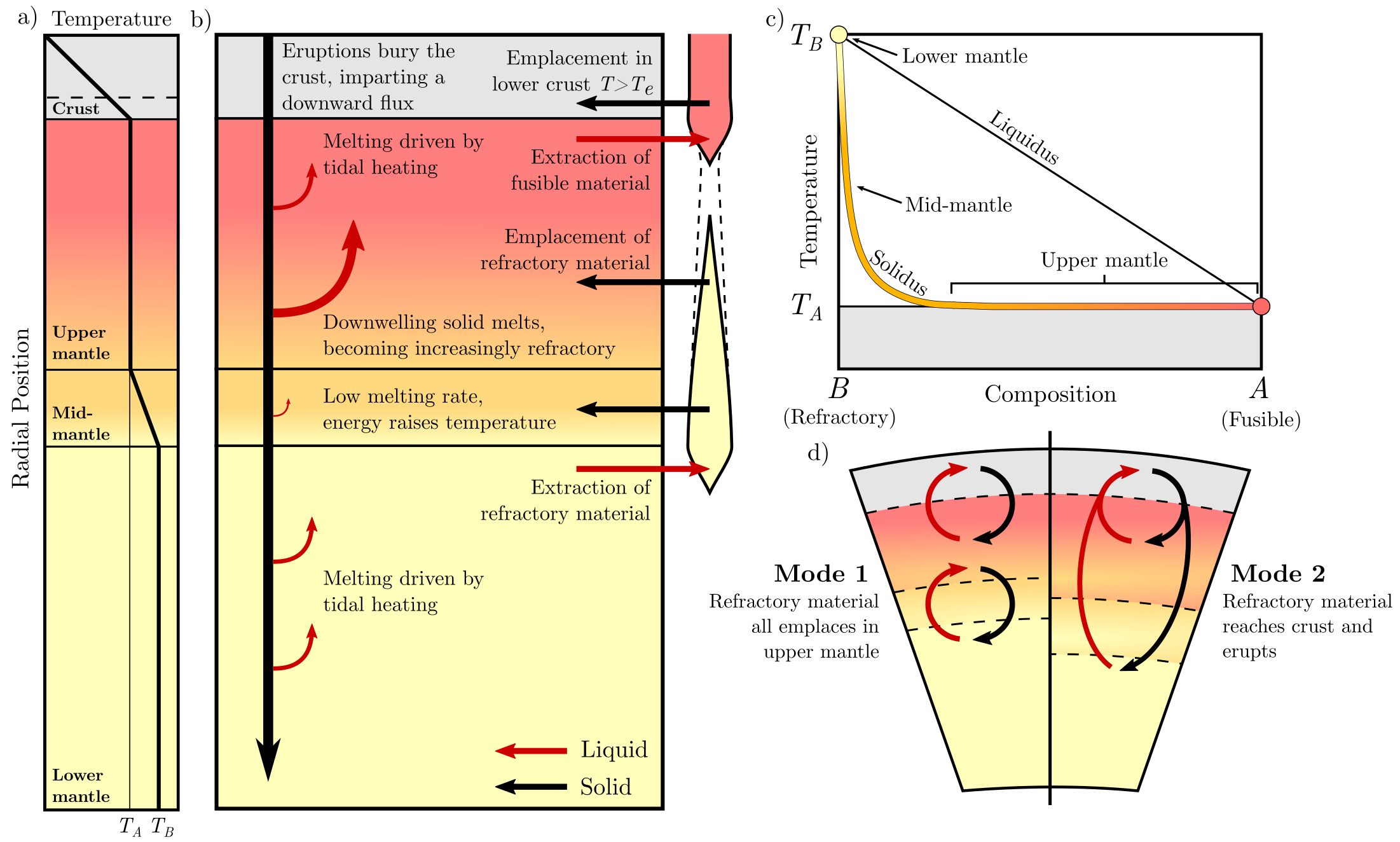}
    \caption{Schematic describing the steady-state solutions. Colour indicates composition (panel c). a) The upper and lower mantle are at the melting point of the fusible and refractory components respectively. b) Melting in the lower mantle is driven by tidal heating. Melting rate in the mid-mantle is low because energy goes toward raising the temperature of downwelling material. If emplacement of refractory melts in the upper mantle occurs, this drives large amount of melting and exhausts the plumbing system material. Fusible melt is extracted from the top of the upper mantle and combines with any plumbing-system material, some of which is emplaced in the lower crust; the remainder rises to fuel volcanic eruptions \cite{spencer_magmatic_2020}. d) In mode 1, all refractory material is emplaced in the upper mantle. In mode 2, refractory material rises to the crust and so cycles through the surface.}
    \label{fig:mass_comp_schematic}
\end{figure}

The two modes share various features that can be identified from figure \ref{fig:SS_example}. We discuss these similarities before considering their differences. Some features are similar to those in the one-component case of \citeA{spencer_magmatic_2020}, which we cover only briefly here. The radial porosity profiles in figure \ref{fig:SS_example}b,f show that the uniform tidal heating causes melt to form throughout the mantle. Figure \ref{fig:SS_example}c,g shows that these melts rise buoyantly while the solid correspondingly sinks. Where melt reaches high pressure it is extracted into the plumbing system, through which it continues to rise. The crustal plumbing system carries melt to the surface where it erupts. The globally-averaged eruption rate is the surface plumbing-system flux in figure \ref{fig:SS_example}c,g. Over long timescales and given the negligible surface conduction, this global eruption rate must extract heat at the same rate that it is input to the interior by tidal heating. The upward flux of melt through the crustal magmatic plumbing system is balanced by downwelling of the solid crust. This recycles erupted material back into the mantle.

At steady state in both modes, the mantle has segregated into three layers: a refractory lower mantle with $T=T_{B}$, a low-melt-fraction mid-mantle with $T_{A}<T<T_{B}$, and a fusible upper mantle with $T\approx T_{A}$. As crustal solid downwells through the upper mantle, tidal heating causes the formation of fusible melts, which buffers the temperature close to $T_{A}$. With continued melting and the buoyant segregation of fusible melts, material downwelling out of the upper mantle is almost exhausted in fusible material and so its solidus temperature has increased according to the phase diagram. In this mid-mantle region, tidal heating primarily acts to raise the temperature of the solid. As a result, melting rate and porosity are low in the mid-mantle, as seen in both modes in figure \ref{fig:SS_example}b,f. Further, the Darcy flux in the mid-mantle is approximately zero (figure \ref{fig:SS_example}c,g), so heat transport across this region occurs only by conduction, advection in the plumbing system, and downward solid advection, a result that we discuss below. Continued heating as the solid downwells through the mid-mantle melts out the remaining small amount of fusible material, and the solid is raised to the refractory melting point $T_{B}$. Melting rate and thus porosity increase in the lower mantle because, as in the upper mantle, all imparted tidal heating directly causes melting.

Magma rising through a two-phase medium cannot pass into impermeable regions. Such regions act as barriers to flow, causing an increase in magma pressure, which forces the solid to decompact and produces higher melt fractions (figure \ref{fig:SS_example}b,f). The crust represents such an impermeable barrier to melts rising through the upper mantle, and similarly, the mid-mantle region acts as an essentially impermeable barrier to melts rising from the lower mantle. The high liquid pressure below these layers causes melts to be extracted into the magmatic plumbing system. Magma extracted from the lower mantle is composed entirely of the refractory component and is at temperature $T_{B}$. Flow through the plumbing system enables these refractory melts to migrate from the lower mantle into the colder overlying mantle and crust. The differences between the two modes are then a consequence of what happens to this melt. The mid- and upper mantle are below the melting point of the refractory component, and it may be expected that these lower temperatures causes refractory plumbing system material to be emplaced during ascent.

In mode 1 (figure \ref{fig:SS_example}a--d), this emplacement is significant --- it acts to exhaust the plumbing system of refractory material before it reaches the crust. As refractory melts are emplaced they release their latent heat to the upper mantle, providing additional heat to melt surrounding fusible material. This is reflected in the rapid increase of Darcy flux in the lower part of the upper mantle in figure \ref{fig:SS_example}c. The emplacement of refractory melts into the upper mantle eventually exhausts the material in the plumbing system, as shown by the plumbing system flux in \ref{fig:SS_example}c. Where the plumbing system material runs out, the melting rate in the upper mantle decreases to just that produced by tidal heating, which causes the change in gradient of the Darcy flux in the upper mantle in figure \ref{fig:SS_example}c. The change in melting rate caused by the cessation of emplacement means that downwelling solid must suddenly decompact, creating a high-porosity decompacting layer in the upper mantle, which can be seen in figure \ref{fig:SS_example}b.

Mode 2 (figure \ref{fig:SS_example}e--h) is the case where at least some of the melt that is extracted from the lower mantle makes it all the way to the surface. The end-member shown in figure \ref{fig:SS_example} is when $h_{M}=0$, in which case there is no emplacement in the upper mantle at all. The plumbing-system flux still decreases in figure \ref{fig:SS_example}g, but only due to radial spreading in a spherical coordinate system, and so the total volume of melt extracted from the lower mantle reaches the top of the upper mantle. Fusible magmas extracted at the top of the upper mantle combine with refractory plumbing system melts rising from below, producing crustal plumbing-system material with a volumetrically averaged temperature and composition. This crustal plumbing-system material describes either an average of non-interacting melts of different temperatures and compositions, or a mixture with an intermediate composition; we assume that the effect is the same on the long timescales considered here. The crustal plumbing system melts are emplaced into the crust at a rate determined by $h_{C}$ and the temperature of the melt, and with a distribution determined by $T_{e}$ \cite{spencer_magmatic_2020}. Material that erupts onto the surface in mode 2 is at a higher temperature than in mode 1, and so serves as a more efficient heat-loss mechanism. This increased heat-loss efficiency results in a lower eruption rate and a thinner crust (see below).

Figure \ref{fig:mass_comp_schematic} shows a schematic of temperature, mass transport, and the phase diagram. Colours in figure \ref{fig:mass_comp_schematic} denote composition according to the phase diagram in panel c. Mode 1 is characterised by a strong segregation of fusible and refractory material; refractory material does not erupt, instead it is cycled between the lower mantle and the deep parts of the upper mantle, whilst fusible material is cycled between the upper mantle and the crust. In mode 2, refractory material is cycled from the lower mantle to the surface, and fusible material is cycled from the upper mantle to the surface. In both modes, the lower mantle is composed purely of refractory material, and the mid-mantle spans compositions corresponding to the steep section of the solidus in figure \ref{fig:mass_comp_schematic}c. In mode 1 there is a transition from almost pure refractory to pure fusible material above the region of the upper mantle where emplacement takes place (figure \ref{fig:SS_example}d). In mode 2, the segregation of the mantle is much less complete, as shown by figure \ref{fig:SS_example}h. The lack of mantle emplacement means that refractory melts rise all the way to the surface. The intermediate-composition erupted material is buried down through the crust and upper mantle, and its composition gradually changes due to the melting of the fusible material by tidal heating.

\subsection{Time-Evolution to Steady State}
\begin{figure}[ht!]
    \centering
    \includegraphics[width=\linewidth]{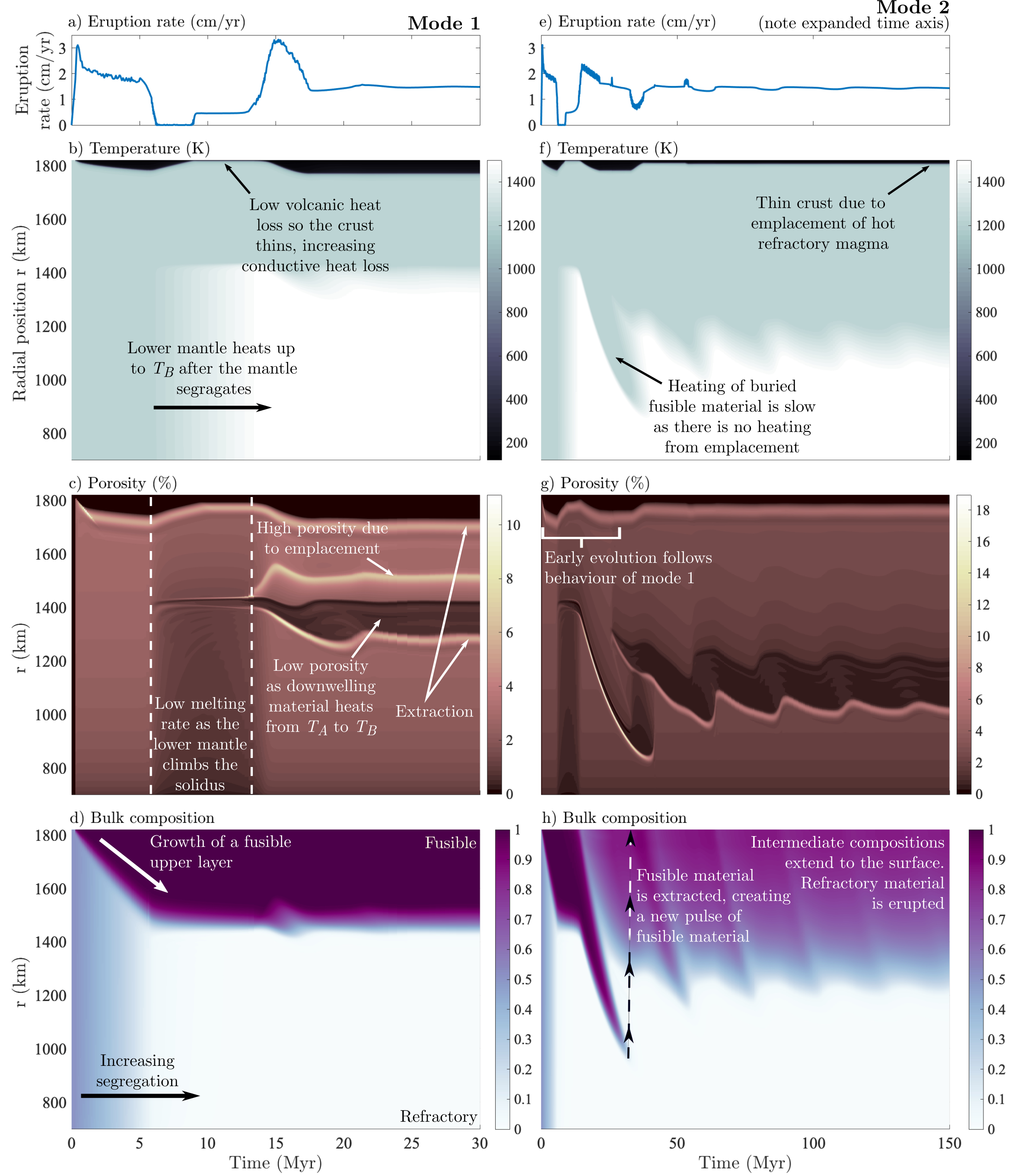}
    \caption{The evolution of the full model to steady state, showing eruption rate, temperature, porosity, and phase-averaged (bulk) composition. Panels a--d show mode 1 of the model where $h_{M}=0.3~$Myr$^{-1}$, and panels e--h show mode 2 where $h_{M}=0$. In both cases, the initial condition is an unstratified mantle of composition $\overline{c}=0.5$, uniformly on the solidus. In mode 1, the emplacement of deep melts into the upper mantle rapidly drives the system to segregate, and equilibrium is reached in $\sim$30~Myr. No refractory material reaches the surface. Mode 2 takes much longer to reach steady state. In mode 2, refractory melt reaches the surface and intermediate compositions exist throughout the upper mantle.}
    \label{fig:t_evolve}
\end{figure}

Figure \ref{fig:t_evolve} shows how both modes of the model evolve to steady state, presenting results for eruption rate, temperature, porosity, and composition. We assume an initially homogeneous body with a bulk composition of $50\%$ fusible material that is initially on its solidus throughout. Other initial conditions, for example starting uniformly cold, or with a cold crust, result in the same broad behaviour, but starting on the solidus removes the spin-up time required to heat the mantle. Thus, despite not knowing the precise `initial condition', various distinctive behaviours can be found that may have important implications for the evolution of Io and other volcanic bodies. The left column of figure \ref{fig:t_evolve} shows the evolution of mode 1, and the right column shows the evolution of mode 2. Note that steady state is reached much more rapidly in mode 1 and so the time axis of mode 2 is significantly expanded. The final steady states are those shown previously in figure \ref{fig:SS_example}.

The early ($t\leq 5~$Myr) evolution of the model is the same for both modes. Fusible (pure-$A$) melts are produced throughout the mantle and rise upward. They are erupted onto the surface and so a cold fusible crust begins to grow. The upper mantle is being continually resupplied with fusible material as it is buried through the crust and remelted at its base. There is no such resupply of fusible material to the deep mantle, which becomes increasingly refractory. After $\sim 5~$Myr, about $20\%$ of Io's volume has been erupted and reburied; the lower mantle is almost completely depleted in fusible material. As a result, melting rate there drops and the solid starts to climb the solidus toward $T=T_{B}$ (figure \ref{fig:eutectic}). Panels a and d in figure \ref{fig:t_evolve} show that the decreased melting rate in the lower mantle reduces the eruption rate to almost zero. This reduction in eruption rate causes the crust to thin, increasing conductive heat loss from the surface. Once the lower mantle has been heated to $T_{B}$, the 3-layer mantle structure described above in the steady-state solution emerges. From this point in the evolution onward, the mid-mantle is acting as an impermeable barrier to refractory melts formed in the lower mantle. The presence of this barrier causes melt to accumulate at the top of the lower mantle, as shown by the bright region at $\sim 1300~$km in figure \ref{fig:t_evolve}c. The accumulation of melt at the top of the lower mantle increases liquid overpressure, which initiates the extraction of refractory melt to the magmatic plumbing system. It is at this point, after around $15~$Myr, that the evolution of the two modes diverge.

In mode 1, the emplacement of the refractory melts into the upper mantle creates a band of intermediate composition there, but the top of the upper mantle and the crust remain purely composed of the fusible material. Steady state is reached after $\sim 30~$Myr, coinciding with the attainment of thermal equilibrium, where heat loss from eruptions equals that input by tidal heating. In mode 2, the deep refractory melts make it to the surface, and the crust --- initially composed of purely fusible material --- becomes of intermediate composition. As there is little to no emplacement in the upper mantle the downwelling crust maintains its composition, which results in cyclic behaviour where the composition of new crust depends on the downwelling composition of the crust a few Myr previously. For example, the initial, purely fusible crust creates a pulse of fusible melt at $\sim 40~$Myr, which produces a new pulse of erupta, more fusible than that in the intervening period. This cycle continues with a decreasing amplitude of differences between erupta compositions until eventually a steady state is reached after around $\sim 200~$Myr. Thermal equilibrium is reached after $\sim 100~$Myr, which can be seen by the constant eruption rate after $\sim 100~$Myr in figure \ref{fig:t_evolve}e.

\subsection{Bulk Composition and Mantle Emplacement Rate}
\begin{figure}[ht!]
    \centering
    \includegraphics[scale=0.2]{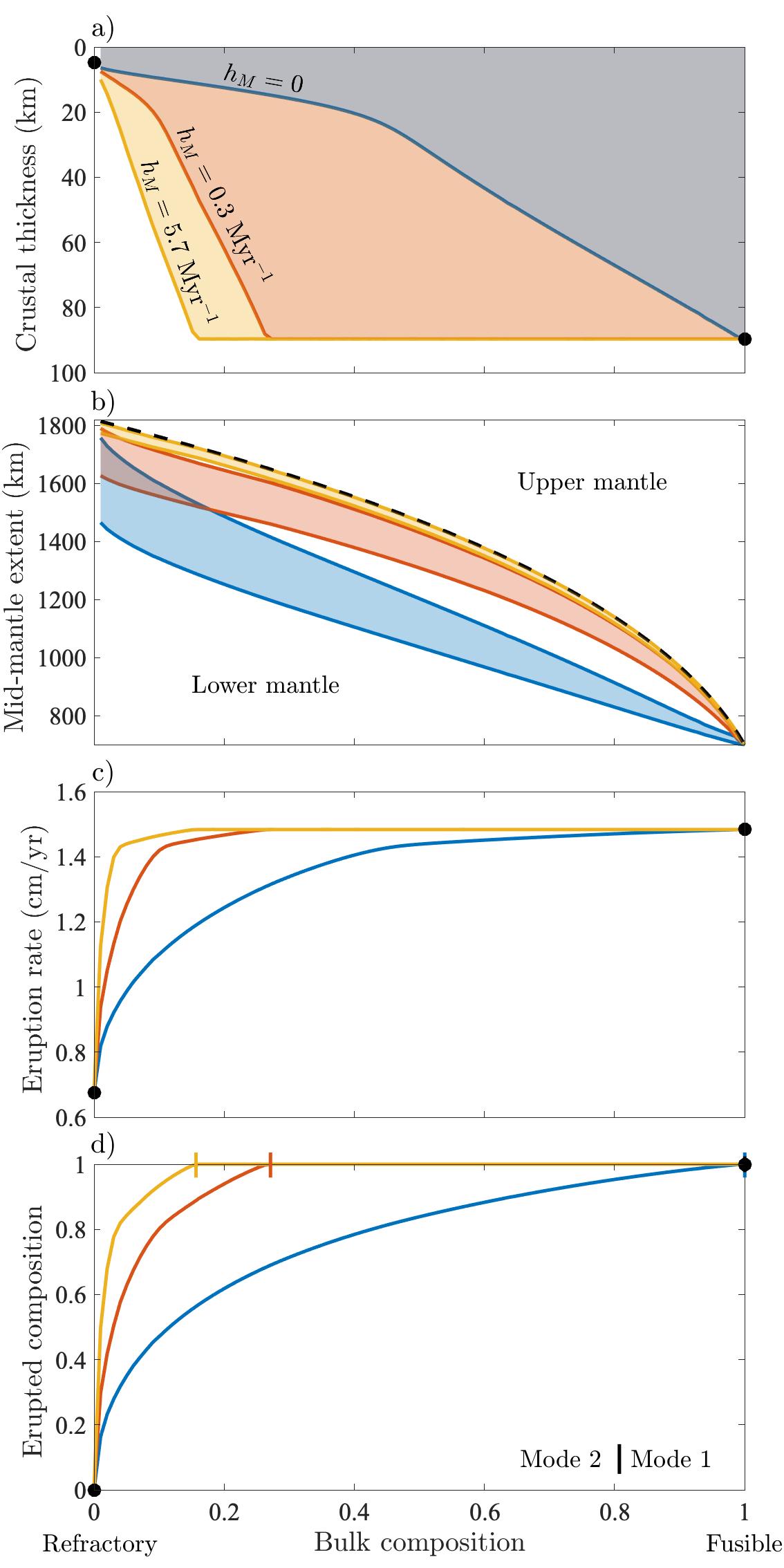}
    \caption{Reduced model solutions of a) crustal thickness, b) location of the mid-mantle, c) eruption rate, and d) erupted composition for varying bulk composition, for three values of $h_{M}$. Refractory material can reach the crust (mode 2) when $h_{M}$ is low, and/or when the bulk composition is refractory (panel d). Higher temperature erupta provides a more efficient heat loss mechanism, so at steady state the eruption rate must decrease (panel c), and this results in a thinner crust (panel a). When refractory material is all frozen in the mantle (at higher values of $h_{M}$ or more fusible bulk compositions), the system is in mode 1. High values of $h_{M}$ create a smaller lower mantle and a large upper mantle for a given bulk composition (panel b). The dotted line on panel b shows the boundary between the upper and lower mantle if fully segregated.}
    \label{fig:cbulk_vary}
\end{figure}

Figure \ref{fig:cbulk_vary} shows how crustal thickness, mantle structure, eruption rate, and erupted composition vary as a function of bulk composition for three values of $h_{M}$. The primary control on whether the model is in mode 1 or mode 2 is the mantle emplacement constant $h_{M}$, but figure \ref{fig:cbulk_vary} shows that bulk composition also exerts a significant control. The results in figure \ref{fig:cbulk_vary} are produced using the reduced steady-state model, which is developed in \ref{appendix:reduced}. The agreement of the reduced model and the full model is demonstrated in figure \ref{fig:SS_example}.

Refractory bulk compositions produce bodies with large refractory lower mantles and thin fusible upper mantles, as shown by figure \ref{fig:cbulk_vary}b. If $h_{M}=0$, all of this refractory material reaches the crust upon melting and the model is always in mode 2. When $h_{M}>0$, some of the refractory material is emplaced and if there is too little of it (i.e., if the bulk composition is fusible enough) then it is all emplaced before reaching the surface and the erupted composition is purely fusible (mode 1). For a given value of $h_{M}$ there is a critical bulk composition that divides mode 2 from mode 1 (figure \ref{fig:cbulk_vary}d). Equivalently, for a given bulk composition there is a critical $h_{M}$ which divides mode 2 (low $h_{M}$) from mode 1 (high $h_{M}$).

A prominent feature of figure \ref{fig:cbulk_vary} is that the crustal thickness and eruption rate both decrease at more refractory bulk compositions. When hot, refractory melt reaches the surface, the eruption rate and crustal thickness drop. The drop in crustal thickness is due to increased emplacement and the higher temperature of the material that is emplaced \cite{spencer_magmatic_2020}. The implications of this are discussed below.

\section{Discussion}
\label{section:discussion}
Our results demonstrate that magmatic segregation and volcanic eruptions lead to a rapid stratification of the mantle. Fusible material is cycled in the upper mantle and crust, and its depletion at depth generates a refractory lower mantle that rises to its melting point. The fate of high-temperature refractory magmas formed in the lower mantle controls the degree of chemical stratification and the composition and temperature of erupted products. If high-temperature refractory melts freeze in the upper mantle (mode 1), no refractory lavas will be observed at the surface and the mantle will be fully stratified. Alternatively, if refractory melts can migrate to the surface (mode 2), refractory eruptions will be observed and the mantle will not be fully stratified.

We first discuss the stratification caused by magmatic segregation and the mantle structure it produces. Next we discuss the key results from each mode, analysing their successes and shortcomings in explaining present observations, and their predictions for future observations. We then consider how lower mantle extraction and the migration of deep refractory melts could be interpreted physically, before finally discussing the limitations and future directions of this work.

\subsection{Stratification by Magmatic Segregation}
The formation of a pure-refractory lower mantle at steady state is a necessary consequence of magmatic segregation in our model. Magmas that form in the lower mantle rise toward the upper mantle, leaving behind an increasingly refractory residuum, a feature shown in the time evolution plots in figure \ref{fig:t_evolve}. The composition of the lower mantle only reaches steady state when all fusible material has been removed. Compositional stratification in our model can be best understood by noting that solids are continually moving downward (see solid flux in figure \ref{fig:SS_example}c,g), and are continually heated as they downwell. Continued heating of intermediate compositions produces fusible melts that segregate buoyantly upward, leading to increasingly refractory compositions with depth.

The structure of the mid- and upper mantle depends on both the phase diagram and the fate of refractory magmas produced in the lower mantle. For our simple two-component phase diagram, the upper mantle is at the fusible melting temperature $T_{A}$, and the mid-mantle must span the temperature range between $T_{A}$ and the temperature $T_{B}$ of the pure-refractory region below. The reduced model, formulated in \ref{appendix:reduced}, shows that the thickness of the mid-mantle ($T_{A}<T<T_{B}$) is determined by the rate at which downwelling solid is heated from $T_{A}$ to $T_{B}$, which is slowest (and thus the mid-mantle is thickest) when no emplacement takes place there. If emplacement of the lower refractory magma there is very efficient (see the largest value of $h_{M}$ and the dotted line in figure \ref{fig:cbulk_vary}b) the mid-mantle is thin and there is almost complete segregation between a pure refractory lower mantle and a pure fusible upper mantle. On the other hand, if refractory melts migrate far into the upper mantle, stratification is less complete. The upward migration reduces the thickness of the pure refractory lower mantle, and increases the thickness of intermediate-composition upper mantle.

With a more detailed phase diagram, we would expect a general structure similar to that proposed here but with greater complexity. In particular, the chemistry of the crust and uppermost mantle would likely be much more complex, with layering controlled by melting temperature, and potentially influenced by near-surface sulphur cycling. Sulphur may be acting as a volatile that reduces melting temperatures \cite{battaglia_ios_2014}. The formation of a lowermost olivine layer is expected to be a feature of any relevent silicate phase diagram, and so our prediction of the formation of high temperature refractory melts is expected to hold. Any temperature range in the mantle over which there is not significant melting would be present as a low-melt-fraction layer that acts as a barrier to melts rising from below, potentially leading to magma overpressure and, in the context of our model, transfer to a plumbing system.

\subsection{Implications of the Two Magmatic Modes}
In this section we discuss the specific results and implications of each mode, analysing the degree to which each mode can explain current observations, and the predictions they make of future observations.

In mode 1, high-temperature refractory magmas formed in the lower mantle migrate into the upper mantle and freeze, delivering their latent heat to the fusible surroundings. The additional melting this emplacement causes can manifest as a high-melt-fraction decompacting layer, as seen in figure \ref{fig:SS_example}b. Magnetic induction models have been used to infer the presence of a $\geq 50~$km region of $\geq 20\%$ melt fraction beneath Io's crust \cite{khurana_evidence_2011}. This has been previously interpreted as a region of concentrated tidal heating \cite{tobie_tidal_2005,bierson_test_2016}, or as a decompacting boundary layer \cite{spencer_magmatic_2020}. Mode 1 of our model shows another manifestation of this decompaction hypothesis; a high melt fraction layer can arise due to freezing of deep refractory melts into the upper mantle. This is a result of the viscous resistance of the mantle to decompaction, and does not occur if the viscosity of the mantle is small, as shown by the solutions to the reduced model in figure \ref{fig:SS_example}, in which this viscous resistance is effectively ignored. A decompacting layer, whether caused by freezing or the strength of the crust \cite{spencer_magmatic_2020}, provides a means of generating high melt fractions in the upper mantle without requiring concentrated tidal heating in this layer.

Mode 1 predicts that no eruptions of refractory material take place. This could be considered consistent with the lack of observed olivine on the surface of Io, although this apparent absence may simply reflect an observational limitation \cite{geissler_global_1999}. The key deviation of mode 1 from observations is that it does not predict any high temperature eruptions. For mode 1 to produce high temperature eruptions would require invoking processes like viscous heating on ascent \cite{keszthelyi_new_2007}.

In mode 2, refractory melts formed in the lower mantle rise to the base of the crust and are ultimately erupted. This predicts the presence of refractory phases on the surface. If Io behaves according to mode 2 of our model, the abundance of refractory phases at the surface could be used to constrain the intrusive behaviour and bulk composition through model outputs like those in figure \ref{fig:cbulk_vary}. The relative lack of upper mantle emplacement in mode 2 means that melting throughout the upper mantle is caused predominantly by tidal heating \cite{moore_thermal_2001,spencer_magmatic_2020}. A key strength of mode 2 in relation to observations lies in its prediction of high eruption temperatures. This provides a means of reconciling heat flow arguments that require heat transport by magmatic segregation \cite{moore_tidal_2003,breuer_10.08_2015}, with observations of high temperature eruptions \cite{mcewen_high-temperature_1998,de_kleer_near-infrared_2014}. Mode 2 supports the hypothesis of \citeA{keszthelyi_magmatic_1997} that eruptions of deep, refractory melts formed within a stratified Io could produce very high temperature lavas. This study expands on that suggestion, demonstrating the dynamical conditions necessary for such eruptions. The rise of deep refractory melts to the surface is a means of recycling deep material to the crust, and so the upper mantle is never fully depleted in refractory material.

The eruption rate predicted by mode 2 is lower than that in mode 1. At steady state and given the negligible surface conduction, the heat lost through eruptions must equal that input by tidal heating \cite{spencer_magmatic_2020}. Increasing the temperature of erupted material means therefore that a lower eruption flux is needed (figure \ref{fig:cbulk_vary}c). Despite this decreased eruption rate, in our model there is very little change in total melting. The combination of the decreased eruption rate and the approximately constant total melt production means that more emplacement of intrusions takes place in bodies operating in mode 2. This effect was explained by \citeA{spencer_magmatic_2020}, where it was shown that the emplaced fraction is given by $C(T_{\textrm{erupt}}-T_{\textrm{surf}})/(L + C(T_{\textrm{erupt}}-T_{\textrm{surf}}))$, where $C$ is the specific heat capacity, $L$ is the latent heat, $T_{\textrm{erupt}}$ is the eruption temperature, and $T_{\textrm{surf}}$ is the surface temperature. The increased emplacement yields a thinner crust than mode 1 for the same value of the crustal emplacement constant $h_{C}$ \cite{spencer_magmatic_2020}. However we note that the appropriate value of $h_{C}$ is not known, so larger crustal thicknesses could also be produced in mode 2, with emplacement spread over a larger region.

A conclusive detection of olivine on Io's surface would provide significant support for mode 2, though we note that processes such as fractional crystallisation may evolve magmas in the crust, meaning that a lack of surface olivine cannot conclusively rule out mode 2. Further, additional observations to constrain the globally averaged volcanic eruption rate and eruption temperature would also test whether refractory melts are migrating out of the deep mantle. On the basis of its ability to explain high eruption temperatures originating from a mantle governed by magmatic segregation, we propose that mode 2 is the more likely state for Io.

\subsection{Mechanism of Ascent for Deep Refractory Magmas}
A fundamental assumption of our model is that deep refractory melts are able to migrate out of the lower mantle without equilibration as they rise. From a modelling perspective, we assume that this occurs due to the accumulation of magmatic overpressure in the lower mantle, which enables melt to leave the lower mantle through some arbitrary `magmatic plumbing system'. In the model, this plumbing system is treated in the same way as the plumbing system in the crust, which we envision as a network of dikes. However, its physical manifestation in the mantle may well be different. In this section we first discuss the assumption that refractory magmas can leave the lower mantle, and then discuss possible physical interpretations of the plumbing system.

If Io is indeed in a thermal steady state \cite{lainey_strong_2009}, heat supplied to the lower mantle must be able to leave to the upper mantle. The heat being transported is primarily in the form of latent heat \cite{moore_thermal_2001}, which can only be lost by the freezing of lower mantle melts. If lower mantle melt was not extracted to a plumbing system, it would have to freeze at the top of the lower mantle where the temperature drops, passing its latent heat to fusible material at the base of the upper mantle, which would melt and continue heat transport upward. We consider such a perfect exchange of mass and energy unlikely due to the extreme liquid overpressures it would generate. We would expect these large liquid overpressures to cause melt to penetrate the overlying upper mantle, which is at its solidus and so is unlikely to have significant strength. Our mantle magmatic plumbing system is intended to capture the range of possible fates of this lower mantle melt. The ultimate freezing and heat transfer could take place at the very base of the upper mantle (large $h_{M}$); in a distributed region of the upper mantle (intermediate $h_{M}$); or in the crust and on the surface (small or zero $h_{M}$).

Assuming then that magma does leave the lower mantle, its rise could be accomplished in a number of ways. The lower mantle is hotter and, at the top, has a higher porosity than the overlying mid- and upper mantle. Together these create a lower bulk density that gives the potential for a Rayleigh-Taylor overturn. In our model, the entire mantle is on its solidus, so we would not expect significant resistance to such an overturn on long timescales. In this interpretation, $h_{M}$ parameterises the equilibration of rising refractory plumes with their surroundings. If the plumes are large and rise rapidly, the degree of equilibration may be very low, representing mode 2 of our model. Such an overturn represents a mode of convective heat transport. Another possibility is that lower mantle melts rise through a system of dikes. High magma pressure in the decompacting boundary layer may localise and nucleate fractures that are driven by magmatic buoyancy. It is possible that such conduits become semi-permanent features, although this would require large amounts of lateral melt transport in the decompacting boundary layer. Interpretations of our deep magmatic plumbing system as a system of dikes would presumably imply a higher value of $h_{M}$ than large convective plumes. Related to the concept of lower mantle melts rising through dikes is the formation of reactive channels. If rising refractory melts are corrosive to more fusible compositions, they can localise into high-flux channels \cite{kelemen_extraction_1995,rees_jones_reaction-infiltration_2018}. Rising lower mantle melts are undersaturated in SiO$_{2}$ and so may dissolve pyroxene and precipitate olivine. This could create high permeability, pure-olivine conduits that allow for the rapid upward rise of refractory melts.

We emphasise that our model makes no explicit assumption about the nature of this plumbing system, other than that it provides some mechanism for upward transport with an efficiency determined by the parameter $h_{M}$. Further work might pursue a more detailed mechanistic interpretation, but that is beyond the current scope.

\subsection{Model Limitations and Future Work}
This work represents an initial step toward a full coupling of geodynamics and thermo-chemistry in volcanic bodies like Io. We have used a simplified phase diagram that, whilst providing useful insight into the general processes of stratification, could be significantly extended. Revisiting previous thermochemical modeling \cite{keszthelyi_magmatic_1997,keszthelyi_new_2007} in light of the dynamics presented here could give a more realistic picture of the compositional structure of Io. The present work also ignores the pressure dependence of melting temperature, the different latent heats of refractory and fusible material, and solid-state phase changes. While we justified these simplifications, a more complete model would aim to incorporate their effects. Further, this work did not consider the possibility that the two chemical components and their melts may have different densities. This, and the one-dimensionality of the model preclude our ability to investigate thermochemical convection, which may be an important part of this system, as discussed below.

In this work we have also neglected the radial distribution of tidal heating. In \citeA{spencer_magmatic_2020} it was demonstrated that the crustal balances of eruption, emplacement and crustal thickness depend only on the integrated heating from below, not its distribution. In the present case, the thicknesses and melt fractions of the different layers in the model would change with variable tidal heating with radius, but the general principles of stratification and melt migration will hold. Future work may aim to couple dynamic models like that presented here with evolving tidal dissipation models.

Another significant simplification in our model is the assumption of spherical symmetry. Tidal heating is a function of not just radius but also latitude and longitude \cite{segatz_tidal_1988,ross_internal_1990}, and may lead to lateral temperature differences on the order of $\sim 100~$K \cite{steinke_tidally_2020}. Such considerations will be key to deciphering the links between interior dissipation and heat transport, and the surface expression of volcanism. If, as speculated above, convective overturn is a mechanism of upward migration of buoyant refractory melts, then future work should include this inherently symmetry-breaking process. The model here is developed to describe leading-order dynamics and compositional evolution; more detailed three-dimensional models are probably needed to facilitate close comparisons to specific surface observations or to make predictions of the surface distribution of eruption products. Such models would be best constrained by more detailed observations of eruptive heat fluxes, temperatures, and petrology.

\subsection{The Possibility of Solid-State Convection}
\label{section:conv}
A potentially significant limitation to our model is its neglect of compositional and thermal density variations. At the pressures relevant for Io's mantle, Fe is expected to preferentially partition into the melt. Such an interpretation of our compositional model might suggest an unstable density stratification with hot, Fe-depleted, refractory material in the lower mantle, and cooler, Fe-enriched, fusible material in the upper mantle \cite{ballmer_reconciling_2017}. Indeed \citeA{keszthelyi_magmatic_1997} proposed that an Fe-rich mid-mantle would form due to the production of fractionated Fe-rich melts. Unstable density stratifications are expected to result in convective instabilities. In this section we therefore discuss the possibility of convective instabilities arising from the chemical structures predicted in this work.

Consider a highly simplified system of two static layers of thickness $b$ separated by a horizontal boundary, where the upper layer (layer 1) has density $\rho_{1}$ and the lower layer (layer 2) has density $\rho_{2}<\rho_{1}$. Both layers have the same viscosity $\eta$. Such a configuration is susceptible to a Rayleigh-Taylor instability. \citeA{turcotte_geodynamics_2014} show that the fastest-growing wavelength of instability is given by $\lambda = 2.568b$, with a growth rate $\tau_{\alpha} = 13.04\eta/(\rho_{1}-\rho_{2})gb$. Taking $b$ to be half the thickness of Io's mantle, we get a wavelength of $\sim 1300~$km; with a viscosity of $\eta = 10^{20}~$Pa\,s and a density difference of $100~$kg/m$^{3}$, the growth rate of the instability is $\sim 200~$kyr. This can be compared to the $\sim 20~$Myr timescale for advection across half the depth of the mantle. This simple calculation indicates that the structure presented in this work may be susceptible to very long-wavelength convective instabilities. Long-wavelength convective overturn would induce the rise of refractory material, potentially affecting the spatial distributions of eruption products.

The applicability of such a calculation to the full system of downwelling solid and buoyantly segregating magma is not immediately clear, especially given the close links between melting, composition, temperature, and density. If convective overturns are able to re-mix the mantle, the drive for compositional convection will be removed. This may lead to episodic behaviour where the mantle becomes increasingly stratified until a convective overturn occurs and resets the compositional structure. Alternatively, convective overturns may sequester Fe at the base of the mantle, removing it from the system considered here. A full analysis of the propensity for thermochemical convection as a consequence of magmatic segregation and volcanism is an interesting avenue of future research. It would require a two-dimensional (at least) model, and potentially a more elaborate petrological parametrisation to assess the partitioning of Fe. It is interesting to note, however, that convection in this system will not necessarily alter heat transport in the way that it does in other systems, since conduction already plays an essentially insignificant role in the one-dimensional structure our model has predicted. Heat transport occurs almost entirely through advection of latent heat by the buoyantly ascending melt, which we might expect to be relatively unaffected by convective motion of the solid. On the other hand, the effect of convection on composition would likely be more significant.

\section{Conclusions}
In this work we have demonstrated that magmatic segregation and volcanic eruptions can rapidly lead to significant compositional stratification of Io's mantle. This stratification produces a refractory lower mantle and a fusible upper mantle and crust. Melting of the refractory lower mantle produces high-temperature melts that must leave the lower mantle in order to facilitate heat loss. The fate of these refractory melts controls the degree of stratification of the mantle and the composition and temperature of erupted lavas. If high-temperature, refractory melts reach the surface, they can provide an explanation of the highest temperature observed eruption, but if they stall in the upper mantle, high temperature eruptions are not predicted. We hypothesise that Io's highest temperature eruptions originate from a deep lower mantle, and that their eruption limits the stratification of the upper mantle. Future observations of the petrology and temperature of eruptions will directly test this hypothesis.

\appendix
\section{Scaled Model}
\label{appendix:scaling}
Here we non-dimensionalise the governing equations of the full model. Much of this process is the same as in appendix A of \citeA{spencer_magmatic_2020}. Dimensional parameters and definitions are given in table \ref{table:parameters1}. Scales and definitions of the non-dimensional parameters are given in table \ref{table:parameters2}. We write, for example, $u=u_{0}\hat{u}$ where $u_{0}$ is the solid velocity scale and $\hat{u}$ is the dimensionless velocity, insert similar expressions for all the variables into the equations, and finally drop the hats on the dimensionless quantities to arrive at a dimensionless model. As in \citeA{spencer_magmatic_2020}, for temperature we write $T=T_{\textrm{surf}}+T_{0}\hat{T}$, but here we take $T_{0}=T_{B}-T_{\textrm{surf}}$, so that a non-dimensional temperature of 1 denotes the melting point of refractory material. We assume spherical symmetry and write all quantities as a function of $r$.

The non-dimensional equation for conservation of mass in the crust--mantle and plumbing system are
\begin{linenomath*}
\begin{equation}
    \frac{1}{r^{2}}\pd{}{r}(r^{2}(u + q) = -E + M,
    \label{eq:bulk_mass_scaled}
\end{equation}
\end{linenomath*}
\begin{linenomath*}
\begin{equation}
    \frac{1}{r^{2}}\pd{(r^{2}q_{p})}{r} = E - M.
    \label{eq:plum_mass_scaled}
\end{equation}
\end{linenomath*}
Conservation of the phase-average composition $\overline{c}$ is
\begin{linenomath*}
\begin{equation}
    \pd{\overline{c}}{t}+\frac{1}{r^{2}}\pd{}{r} \left[r^{2}(\phi_{0}\phi u+q)c_{l}\right] + \frac{1}{r^{2}}\pd{}{r} \left[r^{2}(1-\phi_{0}\phi)uc_{s}\right] = - Ec_{l} + Mc_{p}.
    \label{eq:c_bulk_scaled}
\end{equation}
\end{linenomath*}
Conservation of chemical composition in the plumbing system is
\begin{linenomath*}
\begin{equation}
    \frac{1}{r^{2}}\pd{}{r}(r^{2}q_{p}c_{p}) = Ec_{l} - Mc_{p}.
    \label{eq:c_plum_scaled}
\end{equation}
\end{linenomath*}
Darcy's law and the compaction equation become
\begin{linenomath*}
\begin{subequations}
\begin{gather}
    q = \phi^{n}\left(1-\phi_{0}\phi - \delta \pd{P}{r}\right), \label{eq:darcy_scaled} \\
    \frac{P}{\zeta}+ \frac{1}{r^{2}}\pd{}{r} \left[r^{2}\phi^{n}\left( 1-\phi_{0}\phi - \delta\pd{P}{r}\right)\right] = -E, \label{eq:mom_scaled}
\end{gather}
\end{subequations}
\end{linenomath*}
where $\delta$ is a dimensionless compaction parameter defined in \citeA{spencer_magmatic_2020} and table \ref{table:parameters2}. Conservation of energy becomes
\begin{linenomath*}
\begin{equation}
    \pd{H}{t} + \frac{1}{r^{2}}\pd{}{r}(r^{2}(u+q)T) + \frac{\St}{r^{2}}\pd{}{r}(r^{2}(\phi_{0}\phi u + q)) = \frac{1}{\Pe r^{2}}\pd{}{r}\left( r^{2}\pd{T}{r}\right) + \St \psi + M(T_{p}+\St) - E(T+\St),
    \label{eq:E_scaled}
\end{equation}
\end{linenomath*}
where $\Pe$ is the Peclet number, $\St$ is the Stefan number (table \ref{table:parameters2}), and where bulk enthalpy has been scaled by $T_{0}\rho C$. Conservation of energy in the plumbing system is
\begin{linenomath*}
\begin{equation}
    \frac{1}{r^{2}}\pd{}{r}(r^{2}q_{p}T_{p}) = ET - MT_{p}.
    \label{eq:plum_T_scaled}
\end{equation}
\end{linenomath*}

\begin{table}
\caption{Reference scales and non-dimensional parameters}
\centering
\begin{tabular}{l l l l l}
\hline
Quantity & Symbol & Definition & Preferred Value & Units \\
\hline
Tidal heating scale & $\psi_{0}$ & & $4.2\times 10^{-6}$ & W/m$^{3}$ \\
Liquid velocity scale & $q_{0}$ & $\psi_{0}R/\rho L $ & $6.4\times 10^{-9}$ & m/s \\
Solid velocity scale & $u_{0}$ & $q_{0} $ & $6.4\times 10^{-9}$ & m/s \\
Porosity scale & $\phi_{0}$ & $K_{0}\phi_{0}^{n}\Delta \rho g/\eta_{l}$ & $0.044$ & \\
Temperature scale & $T_{0}$ & $T_{m}-T_{s}$ & $1550$ & K \\
Bulk viscosity scale & $\zeta_{0}$ & $\eta/\phi_{0}$ & $2.3\times 10^{21}$ & Pa\,s \\
Pressure scale & $P_{0}$ & $\zeta_{0}q_{0}/R$ & $8.0\times 10^{6}$ & Pa \\
 & & & & \\
P\'eclet Number & Pe & $q_{0}R/\kappa$ & $1160$ & \\
Stefan Number & St & $L/CT_{0}$ & $0.25$ & \\
Emplacement constant & $\hat{h}$ & $h\rho CT_{0}/\psi_{0}$ &$200$ & \\
Extraction constant & $\hat{\nu}$ & $\nu \zeta_{0}$ & $1000$ & \\
Scaled elastic limit temperature & $\hat{T}_{e}$ & $\frac{T_{e}-T_{s}}{T_{m}-T_{s}}$ & $0.6$ & \\
Compaction parameter & $\delta$ & $\zeta_{0}K_{0}\phi_{0}^{n}/\eta_{l} R^{2} \quad$ & $5.8\times 10^{-3}$ & \\
\hline
\end{tabular}\\
{\raggedright The tidal heating scale $\psi_{0}$ is imposed, which gives the velocity scale $q_{0}$ which in turn gives the porosity scale $\phi_{0}$. \par}
\label{table:parameters2}
\end{table}

\section{Numerical implementation}
\label{appendix:implementation}
Equations \eqref{eq:c_bulk_scaled}, \eqref{eq:c_plum_scaled}, \eqref{eq:mom_scaled}, \eqref{eq:E_scaled}, \eqref{eq:plum_mass_scaled}, and \eqref{eq:plum_T_scaled} are solved for phase averaged composition $\overline{c}$, plumbing system composition $c_{p}$, compaction pressure $P$, enthalpy $H$, plumbing system flux $q_{p}$, and plumbing system temperature $T_{p}$ respectively, using the finite volume method. Other variables are obtained from these six primary variables. In particular enthalpy and phase-averaged composition uniquely define temperature, porosity, solid composition, and liquid composition through the solidus and liquidus equations \eqref{eq:solidus}--\eqref{eq:liquidus}, the scaled definition of bulk enthalpy $H=T+\St \phi_{0}\phi$, and the definition of phase averaged composition $\overline{c}=\phi_{0}\phi c_{l} + (1-\phi_{0}\phi)c_{s}$. This local (cell-wise) problem is solved with a Newton method.

For the numerical solution, we introduce a small amount of artificial diffusion of phase-averaged composition into the system as it helps to avoid discontinuous gradients in composition. The modified composition equation including this artificial diffusion is
\begin{linenomath*}
\begin{equation}
    \pd{\overline{c}}{t}+\frac{1}{r^{2}}\pd{}{r} \left[r^{2}(\phi_{0}\phi u+q)c_{l}\right] + \frac{1}{r^{2}}\pd{}{r} \left[r^{2}(1-\phi_{0}\phi)uc_{s}\right] = \frac{D_{c}}{r^{2}}\pd{}{r}\left(r^{2}\pd{\overline{c}}{r}\right) - Ec_{l} + Mc_{p},
\end{equation}
\end{linenomath*}
where $D_{c}$ is a constant that controls the size of the artificial diffusion. A value of $D_{c}\sim 5\times 10^{-4}$ is generally required for robust convergence, and can be decreased with grid refinement. The effect of this diffusion can be seen in figure \ref{fig:SS_example}d,h where the solid composition of the full model deviates slightly from that of the reduced model. Figure \ref{fig:SS_example} (along with other tests not shown here) shows that the introduction of this diffusion does not affect the model results.

\begin{figure}
    \centering
    \includegraphics[scale=0.15]{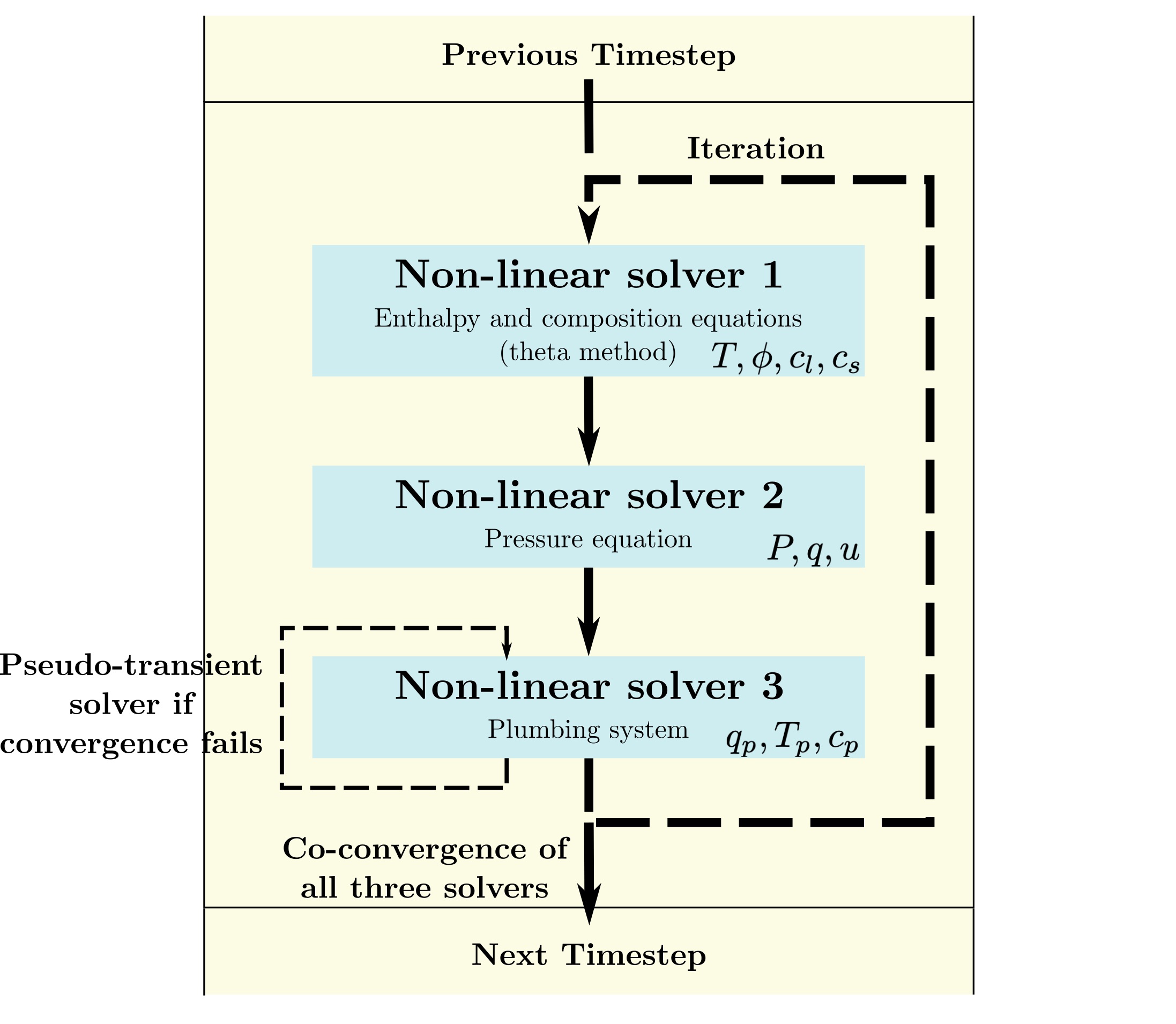}
    \caption{Schematic of the solver used for the full model. The system is split into three non-linear solvers for enthalpy and composition, pressure, and the plumbing system. The solutions to each solver are iterated until all solvers agree to within some small tolerance. A pseudo-transient solver is used for the pipe equations when convergence is poor.}
    \label{fig:solver_schematic}
\end{figure}

The monolithic system (equations \eqref{eq:c_bulk_scaled}--\eqref{eq:plum_T_scaled}) is highly non-linear and tightly coupled \cite{katz_numerical_2007}. Robust convergence is obtained by splitting the system into three non-linear sub-system solvers shown schematically in figure \ref{fig:solver_schematic}. The first sub-system solves equation \eqref{eq:c_bulk_scaled} for phase averaged composition $\overline{c}$, and equation \eqref{eq:E_scaled} for enthalpy $H$. Time integration is performed using the theta method. When $\theta=0$ the system is fully explicit, and is fully implicit when $\theta = 1$. Initially $\theta=0.5$ is used, but if convergence fails an explicit timestep is taken. Sub-system 1 employs Newton's method (with globalization). As part of the residual evaluation for this sub-system, a local non-linear solve for porosity, temperature, and solid and liquid compositions (described above) is required.

Once a solution is found for sub-system 1, the result is passed to solver 2, which solves equation \eqref{eq:mom_scaled} for compaction pressure $P$ using Newton's method (with globalization). This separates the non-linearity of permeability in equation \eqref{eq:mom_scaled} from the composition--enthalpy system in sub-system 1, which also computes porosity. Solver 2 also calculates the Darcy flux $q$ and solid velocity $u$.

Upon convergence, the solutions to the previous two sub-systems are passed to solver 3, which contains the plumbing system equations \eqref{eq:c_plum_scaled}, \eqref{eq:plum_mass_scaled}, and \eqref{eq:plum_T_scaled}. Placing the plumbing system equations in a separate non-linear solver separates them from the pressure dependence of extraction, and the temperature/plumbing system flux dependence of emplacement. Even so convergence can be poor when new regions of extraction emerge, which causes rapid changes to the solutions between timesteps. As per the previous two sub-systems solvers, solver 3 also employs Newton's method (with globalization). If Newton fails to converge, we use a pseudo transient continuation method with implict (backward Euler) time integration. The pseudo transient problem is evolved to steady-state to yield the solutions to equations \eqref{eq:plum_mass_scaled}, \eqref{eq:c_plum_scaled}, and \eqref{eq:plum_T_scaled}.

An adaptive time step is used. At the beginning of each time step $k$, a trial value for the step size $\Delta t_{k} = 1.005~\Delta t_{k-1}$ is selected. The time step is aborted if any of the solvers for the three sub-systems fail to converge, and the step size is reduced by $50\%$. In the event of multiple sub-system solve failures, when $\Delta t_{k} < 1\times 10 ^{-12}$, an explicit timestep is taken using $\Delta t_{k-1}$, and the process of step size reduction is repeated. The simulation is terminated if an explicit step with $\Delta t_{k} < 1\times 10 ^{-12}$ fails to converge.

After the convergence of all three non-linear sub-systems, a unified residual to the monolithic non-linear problem \eqref{eq:c_bulk_scaled}--\eqref{eq:plum_T_scaled} is computed. Successive solution of the three sub-systems are continued until the $\ell_2$-norm of the residual of each discrete PDE is $<1\times 10^{-7}$. Once satisfied, the time step is accepted and the state of the time-dependent PDE is advanced in time from $t_{k}$ to $t_{k+1}=t_{k}+\Delta t_{k}$.

The discretisation and system of non-linear equations is solved using the Portable, Extensible, Toolkit for Scientific computation (PETSc) \cite{petsc-efficient,petsc-web-page,petsc-user-ref}.

\section{Reduced model}
\label{appendix:reduced}
Illuminating simplifications can be made to the full model by assuming small porosity and zero compaction length --- this involves neglecting terms in $\phi_{0}$ and $\delta$ within the scaled equations in \ref{appendix:scaling}. Conservation of composition in the crust-mantle system becomes
\begin{linenomath*}
\begin{equation}
    \pd{\overline{c}}{t}+\frac{1}{r^{2}}\pd{}{r} \left(r^{2}qc_{l}\right) + \frac{1}{r^{2}}\pd{}{r} \left(r^{2}uc_{s}\right) = - Ec_{l} + Mc_{p}.
    \label{eq:reduced_c}
\end{equation}
\end{linenomath*}
We assume that extraction $E$ is zero outside of boundary layers at the base of any solid regions, where it acts to transfer any liquid flux $q$ to the plumbing flux $q_{p}$. $E$ can therefore be thought of as a delta function on the boundaries between partial melt and solid (as boundary layers go to zero thickness in the zero-compaction-length approximation).

Darcy's law and the compaction relation become
\begin{linenomath*}
\begin{subequations}
\begin{gather}
    q=\phi^{n}, \label{eq:darcy_reduced} \\
    \phi P = -\frac{1}{r^{2}}\pd{(r^{2}q)}{r}. \label{eq:comp_rel_red}
\end{gather}
\end{subequations}
\end{linenomath*}
The reduced energy equation \eqref{eq:E_scaled} splits naturally into two cases: `solid', in which case $q=0$ and we have
\begin{linenomath*}
\begin{equation}
    u\pd{T}{r} = \frac{1}{\Pe r^{2}}\pd{}{r}\left(r^{2}\pd{T}{r}\right) + \St \psi + M(T_{p}-T+\St);
    \label{eq:reduced_E_s}
\end{equation}
\end{linenomath*}
and `partially molten', in which case given the phase diagram of pure component $B$ (figure \ref{fig:eutectic}) we have constant $T$ (either at $T_{A}$ or $T_{B}$) and
\begin{linenomath*}
\begin{equation}
    \St\frac{1}{r^{2}}\pd{(r^{2}q)}{r}=\St \psi + M(T_{p}-T+\St),
    \label{eq:reduced_E_PM}
\end{equation}
\end{linenomath*}
where we recall that all extraction occurs on boundaries and so $E$ is absent. In partially-molten regions the compaction pressure is thus given by
\begin{linenomath*}
\begin{equation}
    P = \frac{\St \psi + M(T_{p}-T+\St)}{\St \phi}.
\end{equation}
\end{linenomath*}

Informed by solutions to the full model, we seek solutions that have a partially molten, pure-refractory lower-mantle with $T=T_{B}$ and $\overline{c}=0$, occupying $r_{m}<r<r_{b}$; a mid-mantle solid region $r_{b}<r<r_{a}$ where $T_{A}<T<T_{B}$; an upper-mantle partially molten region $r_{a}<r<r_{c}$ where $T=T_{A}$; and a solid crust $r_{c}<r<R$ where $T_{s}<T<T_{A}$. Note that the mid-mantle region in the full model has non-zero porosity, but since the porosity and Darcy flux there are small, it is treated as a pure solid region in this reduced model.

Throughout, we note that solid velocity $u=-q-q_{p}$ is known from $q$ and $q_{p}$. In the deep refractory mantle, the enthalpy equation \eqref{eq:reduced_E_PM} can be integrated to give
\begin{linenomath*}
\begin{equation}
    q = \frac{\psi}{3}\left( r - \frac{r_{m}^{3}}{r^{2}}\right), \quad \quad r_{m}<r\leq r_{b}.
\end{equation}
\end{linenomath*}
In particular, this gives the value $q_{b}$ at the position $r_{b}$ (which is to be determined). This flux is transferred to the plumbing system, which then has temperature $T_{p}=T_{B}$ and composition $c_{p}=0$. In the region $r_{b}<r<r_{a}$, we have to solve
\begin{linenomath*}
\begin{equation}
    u\pd{T}{r} = \frac{1}{\Pe r^{2}}\pd{}{r}\left(r^{2}\pd{T}{r}\right) + \St \psi + M(T_{p}-T+\St),
    \label{eq:heat_ra}
\end{equation}
\end{linenomath*}
\begin{linenomath*}
\begin{equation}
    \frac{1}{r^{2}}\pd{(r^{2}q_{p})}{r} = -M, \quad \quad M = \hat{h}_{M}(T_{B}-T)\mathcal{I}_{M},
\end{equation}
\end{linenomath*}
where $\mathcal{I}_{M}$ is an indicator function that is zero when $q_{p}=0$ and 1 otherwise. This problem is very similar to that solved for the crust in \citeA{spencer_magmatic_2020}. It is solved with boundary conditions
\begin{linenomath*}
\begin{equation}
\begin{split}
    T=T_{B},\quad \pd{T}{r}=0,\quad q_{p}=q_{b} \quad &\text{at}~r=r_{b} \\
    T=T_{A} \quad &\text{at}~r=r_{a}.
\end{split}
\label{eq:shoot_BC}
\end{equation}
\end{linenomath*}
If the position of $r_{a}$ is known (or guessed --- see below), this problem determines the position of $r_{b}$, as well as the temperature profile and the plumbing flux $q_{p,a}$ at $r_{a}$. This problem can be solved with a shooting method as in \citeA{spencer_magmatic_2020}.

In the partially molten upper mantle ($r_{a}<r<r_{c}$) where $c_{s}\leq 1$, from our phase diagram we have $c_{l}=1$, $c_{p}=0$, and $T=T_{A}$. Equation \eqref{eq:reduced_c} therefore tells us that the solid composition is simply given by
\begin{linenomath*}
\begin{equation}
    c_{s} = -\frac{q}{u}.
\end{equation}
\end{linenomath*}
The emplacement rate $M=\hat{h}_{M}(T_{B}-T_{A})$ is constant and so the plumbing flux is
\begin{linenomath*}
\begin{equation}
    q_{p} = \left(q_{p,a}\frac{r_{a}^{2}}{r^{2}} - \hat{h}_{B}(T_{B}-T_{A})\frac{r^{3}-r_{a}^{2}}{3r^{2}}\right)\mathcal{I}_{qp}
\end{equation}
\end{linenomath*}
where the indicator function $\mathcal{I}_{qp}$ indicates that this quantity cannot go below zero. The reduced enthalpy equation \eqref{eq:reduced_E_PM} then gives
\begin{linenomath*}
\begin{equation}
    q = \frac{\psi}{3}\left(r - \frac{r_{a}^{3}}{r^{2}} \right) + \left( 1+\frac{T_{B}-T_{A}}{\St} \right)\left[ q_{p,a}\frac{r_{a}^{2}}{r^{2}} - q_{p}\right] + q_{a}\frac{r_{a}^{2}}{r^{2}}.
\end{equation}
\end{linenomath*}
The second term here is the melting due to the heat released when material is emplaced from the plumbing system. The final term comes from balancing energy at the interface $r=r_{a}$; since there is a temperature gradient below, the Stefan condition (jump condition for the enthalpy equation) gives a sudden melt flux
\begin{linenomath*}
\begin{equation}
    q_{a} = -\frac{1}{\St\Pe}\pd{T}{r}\bigg|_{-},
\end{equation}
\end{linenomath*}
where the temperature gradient here is known from the solution of \eqref{eq:heat_ra}--\eqref{eq:shoot_BC}. From these solutions we know the plumbing flux $q_{p,c}$ and liquid flux $q_{c}$ arriving at the crust mantle boundary $r_{c}$ (which is to be determined). Since the flux $q_{c}$ is then transferred to the plumbing system, the plumbing system in the crust subsequently has constant temperature and composition given by
\begin{linenomath*}
\begin{equation}
    c_{p}=\frac{q_{c}}{q_{p,c}+q_{c}},\quad\quad T_{p} = \frac{q_{p,c}T_{B}+q_{c}T_{A}}{q_{p,c}+q_{c}}.
\end{equation}
\end{linenomath*}
Note that if all refractory material has been emplaced beneath the crust, then $q_{p,c}=0$ and this simply says that the crustal plumbing system has $c_{p}=1$, and $T_{p}=T_{A}$. Within the region $r_{c}<r<R$, we have to solve the system
\begin{linenomath*}
\begin{equation}
    u\pd{T}{r} = \frac{1}{\Pe r^{2}}\pd{}{r}\left(r^{2}\pd{T}{r}\right) + \St \psi + M(T_{p}-T+\St),
\end{equation}
\end{linenomath*}
\begin{linenomath*}
\begin{equation}
    \frac{1}{r^{2}}\pd{(r^{2}q_{p})}{r} = -M, \quad \quad M = \hat{h}_{C}(T_{p}-T)\mathcal{I}_{M}.
\end{equation}
\end{linenomath*}
This system has the boundary conditions
\begin{linenomath*}
\begin{equation}
\begin{split}
    T=T_{A},\quad \pd{T}{r}=0,\quad q_{p}=q_{c}+q_{p,c},\quad &\text{at}~r=r_{c}, \\
    T=T_{s}\quad &\text{at}~r=R.
\end{split}
\end{equation}
\end{linenomath*}
This system is solved the same way as the mid-mantle solid region: a shooting method is used to find the position $r_{c}$, as well as the crustal temperature distribution and the plumbing flux. Seeking a particular bulk composition for silicate Io, a guess can be made of $r_{a}$, and a Newton method used on the resultant bulk composition to find the position of $r_{a}$ that gives the desired bulk composition.

Figure \ref{fig:SS_example} shows solutions to the reduced model as dashed lines, showing good agreement with the full model. There are slight differences in the position of the mid-mantle that arise in the full model due to the smoothed solidus (equation \eqref{eq:solidus}).

%
%
%
%
%
%
%
%

\acknowledgments
This work was funded by the \textit{Science and Technologies Facilities Council}, the University of Oxford's \textit{Oxford Radcliffe Scholarship}, and \textit{University College, Oxford}. This research received funding from the European Research Council under the European Union’s Horizon 2020 research and innovation programme grant agreement number 772255. DAM acknowledges financial support from the Alfred P. Sloan Foundation through the Deep Carbon Observatory (DCO) Modelling and Visualization Forum. Source code and data used in the production of figures can be found at https://zenodo.org/record/3898245 \cite{spencer_spencer-spaceio_composition_2020}.


%
%

\bibliography{Oxford,petsc_bib}

\begin{thebibliography}{}

\bibitem [\protect \citeauthoryear {%
Balay%
\ \protect \BOthers {.}}{%
Balay%
\ \protect \BOthers {.}}{%
{\protect \APACyear {2019}}%
}]{%
petsc-web-page}
\APACinsertmetastar {%
petsc-web-page}%
\begin{APACrefauthors}%
Balay, S.%
, Abhyankar, S.%
, Adams, M\BPBI F.%
, Brown, J.%
, Brune, P.%
, Buschelman, K.%
\BDBL {}Zhang, H.%
\end{APACrefauthors}%
\unskip\
\newblock
\APACrefYearMonthDay{2019}{}{}.
\newblock
\APACrefbtitle {{PETS}c Web page.} {{PETS}c web page.}
\newblock
\APAChowpublished {https://www.mcs.anl.gov/petsc}.
\newblock
\begin{APACrefURL} \url{https://www.mcs.anl.gov/petsc} \end{APACrefURL}
\PrintBackRefs{\CurrentBib}

\bibitem [\protect \citeauthoryear {%
Balay%
\ \protect \BOthers {.}}{%
Balay%
\ \protect \BOthers {.}}{%
{\protect \APACyear {2020}}%
}]{%
petsc-user-ref}
\APACinsertmetastar {%
petsc-user-ref}%
\begin{APACrefauthors}%
Balay, S.%
, Abhyankar, S.%
, Adams, M\BPBI F.%
, Brown, J.%
, Brune, P.%
, Buschelman, K.%
\BDBL {}Zhang, H.%
\end{APACrefauthors}%
\unskip\
\newblock
\APACrefYearMonthDay{2020}{}{}.
\newblock
\APACrefbtitle {{PETS}c Users Manual} {{PETS}c users manual}\
  \APACbVolEdTR{}{\BTR{}\ \BNUM\ ANL-95/11 - Revision 3.13}.
\newblock
\APACaddressInstitution{}{Argonne National Laboratory}.
\newblock
\begin{APACrefURL} \url{https://www.mcs.anl.gov/petsc} \end{APACrefURL}
\PrintBackRefs{\CurrentBib}

\bibitem [\protect \citeauthoryear {%
Balay%
, Gropp%
, McInnes%
\BCBL {}\ \BBA {} Smith%
}{%
Balay%
\ \protect \BOthers {.}}{%
{\protect \APACyear {1997}}%
}]{%
petsc-efficient}
\APACinsertmetastar {%
petsc-efficient}%
\begin{APACrefauthors}%
Balay, S.%
, Gropp, W\BPBI D.%
, McInnes, L\BPBI C.%
\BCBL {}\ \BBA {} Smith, B\BPBI F.%
\end{APACrefauthors}%
\unskip\
\newblock
\APACrefYearMonthDay{1997}{}{}.
\newblock
{\BBOQ}\APACrefatitle {Efficient Management of Parallelism in Object Oriented
  Numerical Software Libraries} {Efficient management of parallelism in object
  oriented numerical software libraries}.{\BBCQ}
\newblock
\BIn{} E.~Arge, A\BPBI M.~Bruaset\BCBL {}\ \BBA {} H\BPBI P.~Langtangen\
  (\BEDS), \APACrefbtitle {Modern Software Tools in Scientific Computing}
  {Modern software tools in scientific computing}\ (\BPGS\ 163--202).
\newblock
\APACaddressPublisher{}{Birkh{\"{a}}user Press}.
\PrintBackRefs{\CurrentBib}

\bibitem [\protect \citeauthoryear {%
Ballmer%
, Lourenço%
, Hirose%
, Caracas%
\BCBL {}\ \BBA {} Nomura%
}{%
Ballmer%
\ \protect \BOthers {.}}{%
{\protect \APACyear {2017}}%
}]{%
ballmer_reconciling_2017}
\APACinsertmetastar {%
ballmer_reconciling_2017}%
\begin{APACrefauthors}%
Ballmer, M\BPBI D.%
, Lourenço, D\BPBI L.%
, Hirose, K.%
, Caracas, R.%
\BCBL {}\ \BBA {} Nomura, R.%
\end{APACrefauthors}%
\unskip\
\newblock
\APACrefYearMonthDay{2017}{}{}.
\newblock
{\BBOQ}\APACrefatitle {Reconciling magma-ocean crystallization models with the
  present-day structure of the {Earth}'s mantle} {Reconciling magma-ocean
  crystallization models with the present-day structure of the {Earth}'s
  mantle}.{\BBCQ}
\newblock
\APACjournalVolNumPages{Geochemistry, Geophysics,
  Geosystems}{18}{7}{2785--2806}.
\newblock
\begin{APACrefURL}
  [{2020-07-28}]\url{https://agupubs.onlinelibrary.wiley.com/doi/abs/10.1002/2017GC006917}
  \end{APACrefURL}
\newblock
\APACrefnote{\_eprint:
  https://agupubs.onlinelibrary.wiley.com/doi/pdf/10.1002/2017GC006917}
\newblock
\begin{APACrefDOI} \doi{10.1002/2017GC006917} \end{APACrefDOI}
\PrintBackRefs{\CurrentBib}

\bibitem [\protect \citeauthoryear {%
Battaglia%
, Stewart%
\BCBL {}\ \BBA {} Kieffer%
}{%
Battaglia%
\ \protect \BOthers {.}}{%
{\protect \APACyear {2014}}%
}]{%
battaglia_ios_2014}
\APACinsertmetastar {%
battaglia_ios_2014}%
\begin{APACrefauthors}%
Battaglia, S\BPBI M.%
, Stewart, M\BPBI A.%
\BCBL {}\ \BBA {} Kieffer, S\BPBI W.%
\end{APACrefauthors}%
\unskip\
\newblock
\APACrefYearMonthDay{2014}{{\APACmonth{06}}}{}.
\newblock
{\BBOQ}\APACrefatitle {Io’s theothermal (sulfur) – {Lithosphere} cycle
  inferred from sulfur solubility modeling of {Pele}’s magma supply} {Io’s
  theothermal (sulfur) – {Lithosphere} cycle inferred from sulfur solubility
  modeling of {Pele}’s magma supply}.{\BBCQ}
\newblock
\APACjournalVolNumPages{Icarus}{235}{}{123--129}.
\newblock
\begin{APACrefURL}
  [{2019-08-29}]\url{http://www.sciencedirect.com/science/article/pii/S0019103514001456}
  \end{APACrefURL}
\newblock
\begin{APACrefDOI} \doi{10.1016/j.icarus.2014.03.019} \end{APACrefDOI}
\PrintBackRefs{\CurrentBib}

\bibitem [\protect \citeauthoryear {%
Bierson%
\ \BBA {} Nimmo%
}{%
Bierson%
\ \BBA {} Nimmo%
}{%
{\protect \APACyear {2016}}%
}]{%
bierson_test_2016}
\APACinsertmetastar {%
bierson_test_2016}%
\begin{APACrefauthors}%
Bierson, C\BPBI J.%
\BCBT {}\ \BBA {} Nimmo, F.%
\end{APACrefauthors}%
\unskip\
\newblock
\APACrefYearMonthDay{2016}{{\APACmonth{11}}}{}.
\newblock
{\BBOQ}\APACrefatitle {A test for {Io}'s magma ocean: {Modeling} tidal
  dissipation with a partially molten mantle} {A test for {Io}'s magma ocean:
  {Modeling} tidal dissipation with a partially molten mantle}.{\BBCQ}
\newblock
\APACjournalVolNumPages{Journal of Geophysical Research:
  Planets}{121}{11}{2211--2224}.
\newblock
\begin{APACrefURL}
  \url{http://onlinelibrary.wiley.com/doi/10.1002/2016JE005005/abstract}
  \end{APACrefURL}
\newblock
\begin{APACrefDOI} \doi{10.1002/2016JE005005} \end{APACrefDOI}
\PrintBackRefs{\CurrentBib}

\bibitem [\protect \citeauthoryear {%
Bl\"ocker%
, Saur%
, Roth%
\BCBL {}\ \BBA {} Strobel%
}{%
Bl\"ocker%
\ \protect \BOthers {.}}{%
{\protect \APACyear {2018}}%
}]{%
blocker_mhd_2018}
\APACinsertmetastar {%
blocker_mhd_2018}%
\begin{APACrefauthors}%
Bl\"ocker, A.%
, Saur, J.%
, Roth, L.%
\BCBL {}\ \BBA {} Strobel, D\BPBI F.%
\end{APACrefauthors}%
\unskip\
\newblock
\APACrefYearMonthDay{2018}{}{}.
\newblock
{\BBOQ}\APACrefatitle {{MHD} {Modeling} of the {Plasma} {Interaction} {With}
  {Io}'s {Asymmetric} {Atmosphere}} {{MHD} {Modeling} of the {Plasma}
  {Interaction} {With} {Io}'s {Asymmetric} {Atmosphere}}.{\BBCQ}
\newblock
\APACjournalVolNumPages{Journal of Geophysical Research: Space
  Physics}{123}{11}{9286--9311}.
\newblock
\begin{APACrefURL}
  [{2019-11-11}]\url{https://agupubs.onlinelibrary.wiley.com/doi/abs/10.1029/2018JA025747}
  \end{APACrefURL}
\newblock
\begin{APACrefDOI} \doi{10.1029/2018JA025747} \end{APACrefDOI}
\PrintBackRefs{\CurrentBib}

\bibitem [\protect \citeauthoryear {%
Breuer%
\ \BBA {} Moore%
}{%
Breuer%
\ \BBA {} Moore%
}{%
{\protect \APACyear {2015}}%
}]{%
breuer_10.08_2015}
\APACinsertmetastar {%
breuer_10.08_2015}%
\begin{APACrefauthors}%
Breuer, D.%
\BCBT {}\ \BBA {} Moore, W\BPBI B.%
\end{APACrefauthors}%
\unskip\
\newblock
\APACrefYearMonthDay{2015}{{\APACmonth{01}}}{}.
\newblock
{\BBOQ}\APACrefatitle {10.08 - {Dynamics} and {Thermal} {History} of the
  {Terrestrial} {Planets}, the {Moon}, and {Io}} {10.08 - {Dynamics} and
  {Thermal} {History} of the {Terrestrial} {Planets}, the {Moon}, and
  {Io}}.{\BBCQ}
\newblock
\BIn{} G.~Schubert\ (\BED), \APACrefbtitle {Treatise on {Geophysics} ({Second}
  {Edition})} {Treatise on {Geophysics} ({Second} {Edition})}\ (\BPGS\
  255--305).
\newblock
\APACaddressPublisher{Oxford}{Elsevier}.
\newblock
\begin{APACrefURL}
  [{2019-11-11}]\url{http://www.sciencedirect.com/science/article/pii/B9780444538024001731}
  \end{APACrefURL}
\newblock
\begin{APACrefDOI} \doi{10.1016/B978-0-444-53802-4.00173-1} \end{APACrefDOI}
\PrintBackRefs{\CurrentBib}

\bibitem [\protect \citeauthoryear {%
Davies%
\ \protect \BOthers {.}}{%
Davies%
\ \protect \BOthers {.}}{%
{\protect \APACyear {2017}}%
}]{%
davies_novel_2017}
\APACinsertmetastar {%
davies_novel_2017}%
\begin{APACrefauthors}%
Davies, A\BPBI G.%
, Gunapala, S.%
, Soibel, A.%
, Ting, D.%
, Rafol, S.%
, Blackwell, M.%
\BDBL {}Kelly, M.%
\end{APACrefauthors}%
\unskip\
\newblock
\APACrefYearMonthDay{2017}{{\APACmonth{09}}}{}.
\newblock
{\BBOQ}\APACrefatitle {A novel technology for measuring the eruption
  temperature of silicate lavas with remote sensing: {Application} to {Io} and
  other planets} {A novel technology for measuring the eruption temperature of
  silicate lavas with remote sensing: {Application} to {Io} and other
  planets}.{\BBCQ}
\newblock
\APACjournalVolNumPages{Journal of Volcanology and Geothermal
  Research}{343}{}{1--16}.
\newblock
\begin{APACrefURL}
  [{2019-08-29}]\url{http://www.sciencedirect.com/science/article/pii/S0377027317301270}
  \end{APACrefURL}
\newblock
\begin{APACrefDOI} \doi{10.1016/j.jvolgeores.2017.04.016} \end{APACrefDOI}
\PrintBackRefs{\CurrentBib}

\bibitem [\protect \citeauthoryear {%
Davies%
, Keszthelyi%
\BCBL {}\ \BBA {} McEwen%
}{%
Davies%
\ \protect \BOthers {.}}{%
{\protect \APACyear {2016}}%
}]{%
davies_determination_2016}
\APACinsertmetastar {%
davies_determination_2016}%
\begin{APACrefauthors}%
Davies, A\BPBI G.%
, Keszthelyi, L\BPBI P.%
\BCBL {}\ \BBA {} McEwen, A\BPBI S.%
\end{APACrefauthors}%
\unskip\
\newblock
\APACrefYearMonthDay{2016}{{\APACmonth{11}}}{}.
\newblock
{\BBOQ}\APACrefatitle {Determination of eruption temperature of {Io}'s lavas
  using lava tube skylights} {Determination of eruption temperature of {Io}'s
  lavas using lava tube skylights}.{\BBCQ}
\newblock
\APACjournalVolNumPages{Icarus}{278}{}{266--278}.
\newblock
\begin{APACrefURL}
  [{2019-08-29}]\url{http://www.sciencedirect.com/science/article/pii/S0019103516302640}
  \end{APACrefURL}
\newblock
\begin{APACrefDOI} \doi{10.1016/j.icarus.2016.06.003} \end{APACrefDOI}
\PrintBackRefs{\CurrentBib}

\bibitem [\protect \citeauthoryear {%
de Kleer%
, de Pater%
, Davies%
\BCBL {}\ \BBA {} Ádámkovics%
}{%
de Kleer%
\ \protect \BOthers {.}}{%
{\protect \APACyear {2014}}%
}]{%
de_kleer_near-infrared_2014}
\APACinsertmetastar {%
de_kleer_near-infrared_2014}%
\begin{APACrefauthors}%
de Kleer, K.%
, de Pater, I.%
, Davies, A\BPBI G.%
\BCBL {}\ \BBA {} Ádámkovics, M.%
\end{APACrefauthors}%
\unskip\
\newblock
\APACrefYearMonthDay{2014}{{\APACmonth{11}}}{}.
\newblock
{\BBOQ}\APACrefatitle {Near-infrared monitoring of {Io} and detection of a
  violent outburst on 29 {August} 2013} {Near-infrared monitoring of {Io} and
  detection of a violent outburst on 29 {August} 2013}.{\BBCQ}
\newblock
\APACjournalVolNumPages{Icarus}{242}{}{352--364}.
\newblock
\begin{APACrefURL}
  [{2020-04-01}]\url{http://www.sciencedirect.com/science/article/pii/S0019103514003170}
  \end{APACrefURL}
\newblock
\begin{APACrefDOI} \doi{10.1016/j.icarus.2014.06.006} \end{APACrefDOI}
\PrintBackRefs{\CurrentBib}

\bibitem [\protect \citeauthoryear {%
de Kleer%
, de Pater%
\BCBL {}\ \protect \BOthers {.}}{%
de Kleer%
, de Pater%
\BCBL {}\ \protect \BOthers {.}}{%
{\protect \APACyear {2019}}%
}]{%
de_kleer_ios_2019}
\APACinsertmetastar {%
de_kleer_ios_2019}%
\begin{APACrefauthors}%
de Kleer, K.%
, de Pater, I.%
, Molter, E\BPBI M.%
, Banks, E.%
, Davies, A\BPBI G.%
, Alvarez, C.%
\BDBL {}Tollefson, J.%
\end{APACrefauthors}%
\unskip\
\newblock
\APACrefYearMonthDay{2019}{{\APACmonth{06}}}{}.
\newblock
{\BBOQ}\APACrefatitle {Io's {Volcanic} {Activity} from {Time} {Domain}
  {Adaptive} {Optics} {Observations}: 2013–2018} {Io's {Volcanic} {Activity}
  from {Time} {Domain} {Adaptive} {Optics} {Observations}: 2013–2018}.{\BBCQ}
\newblock
\APACjournalVolNumPages{The Astronomical Journal}{158}{1}{29}.
\newblock
\begin{APACrefURL}
  [{2020-06-01}]\url{https://doi.org/10.3847%2F1538-3881%2Fab2380}
  \end{APACrefURL}
\newblock
\APACrefnote{Publisher: American Astronomical Society}
\newblock
\begin{APACrefDOI} \doi{10.3847/1538-3881/ab2380} \end{APACrefDOI}
\PrintBackRefs{\CurrentBib}

\bibitem [\protect \citeauthoryear {%
de Kleer%
, McEwen%
\BCBL {}\ \BBA {} Park%
}{%
de Kleer%
, McEwen%
\BCBL {}\ \BBA {} Park%
}{%
{\protect \APACyear {2019}}%
}]{%
de_kleer_tidal_2019}
\APACinsertmetastar {%
de_kleer_tidal_2019}%
\begin{APACrefauthors}%
de Kleer, K.%
, McEwen, A\BPBI S.%
\BCBL {}\ \BBA {} Park, R.%
\end{APACrefauthors}%
\unskip\
\newblock
\APACrefYearMonthDay{2019}{{\APACmonth{06}}}{}.
\newblock
{\BBOQ}\APACrefatitle {Tidal {Heating}: {Lessons} from {Io} and the {Jovian}
  {System}} {Tidal {Heating}: {Lessons} from {Io} and the {Jovian}
  {System}}.{\BBCQ}
\newblock
\BIn{} \APACrefbtitle {Final {Report} for the {Keck} {Institute} for {Space}
  {Studies}.} {Final {Report} for the {Keck} {Institute} for {Space}
  {Studies}.}
\newblock
\begin{APACrefURL}
  \url{https://www.kiss.caltech.edu/final_reports/Tidal_Heating_final_report.pdf}
  \end{APACrefURL}
\PrintBackRefs{\CurrentBib}

\bibitem [\protect \citeauthoryear {%
de Kleer%
, Nimmo%
\BCBL {}\ \BBA {} Kite%
}{%
de Kleer%
, Nimmo%
\BCBL {}\ \BBA {} Kite%
}{%
{\protect \APACyear {2019}}%
}]{%
de_kleer_variability_2019}
\APACinsertmetastar {%
de_kleer_variability_2019}%
\begin{APACrefauthors}%
de Kleer, K.%
, Nimmo, F.%
\BCBL {}\ \BBA {} Kite, E.%
\end{APACrefauthors}%
\unskip\
\newblock
\APACrefYearMonthDay{2019}{}{}.
\newblock
{\BBOQ}\APACrefatitle {Variability in {Io}'s {Volcanism} on {Timescales} of
  {Periodic} {Orbital} {Changes}} {Variability in {Io}'s {Volcanism} on
  {Timescales} of {Periodic} {Orbital} {Changes}}.{\BBCQ}
\newblock
\APACjournalVolNumPages{Geophysical Research Letters}{46}{12}{6327--6332}.
\newblock
\begin{APACrefURL}
  [{2020-06-01}]\url{https://agupubs.onlinelibrary.wiley.com/doi/abs/10.1029/2019GL082691}
  \end{APACrefURL}
\newblock
\APACrefnote{\_eprint:
  https://agupubs.onlinelibrary.wiley.com/doi/pdf/10.1029/2019GL082691}
\newblock
\begin{APACrefDOI} \doi{10.1029/2019GL082691} \end{APACrefDOI}
\PrintBackRefs{\CurrentBib}

\bibitem [\protect \citeauthoryear {%
Geissler%
\ \protect \BOthers {.}}{%
Geissler%
\ \protect \BOthers {.}}{%
{\protect \APACyear {1999}}%
}]{%
geissler_global_1999}
\APACinsertmetastar {%
geissler_global_1999}%
\begin{APACrefauthors}%
Geissler, P\BPBI E.%
, McEwen, A\BPBI S.%
, Keszthelyi, L.%
, Lopes-Gautier, R.%
, Granahan, J.%
\BCBL {}\ \BBA {} Simonelli, D\BPBI P.%
\end{APACrefauthors}%
\unskip\
\newblock
\APACrefYearMonthDay{1999}{{\APACmonth{08}}}{}.
\newblock
{\BBOQ}\APACrefatitle {Global {Color} {Variations} on {Io}} {Global {Color}
  {Variations} on {Io}}.{\BBCQ}
\newblock
\APACjournalVolNumPages{Icarus}{140}{2}{265--282}.
\newblock
\begin{APACrefURL}
  [{2020-06-01}]\url{http://www.sciencedirect.com/science/article/pii/S0019103599961286}
  \end{APACrefURL}
\newblock
\begin{APACrefDOI} \doi{10.1006/icar.1999.6128} \end{APACrefDOI}
\PrintBackRefs{\CurrentBib}

\bibitem [\protect \citeauthoryear {%
Hay%
, Trinh%
\BCBL {}\ \BBA {} Matsuyama%
}{%
Hay%
\ \protect \BOthers {.}}{%
{\protect \APACyear {2020}}%
}]{%
hay_powering_2020}
\APACinsertmetastar {%
hay_powering_2020}%
\begin{APACrefauthors}%
Hay, H\BPBI C\BPBI F\BPBI C.%
, Trinh, A.%
\BCBL {}\ \BBA {} Matsuyama, I.%
\end{APACrefauthors}%
\unskip\
\newblock
\APACrefYearMonthDay{2020}{}{}.
\newblock
{\BBOQ}\APACrefatitle {Powering the {Galilean} {Satellites} with {Moon}-moon
  {Tides}} {Powering the {Galilean} {Satellites} with {Moon}-moon
  {Tides}}.{\BBCQ}
\newblock
\APACjournalVolNumPages{Geophysical Research Letters}{n/a}{n/a}{e2020GL088317}.
\newblock
\begin{APACrefURL}
  [{2020-07-30}]\url{https://agupubs.onlinelibrary.wiley.com/doi/abs/10.1029/2020GL088317}
  \end{APACrefURL}
\newblock
\APACrefnote{\_eprint:
  https://agupubs.onlinelibrary.wiley.com/doi/pdf/10.1029/2020GL088317}
\newblock
\begin{APACrefDOI} \doi{10.1029/2020GL088317} \end{APACrefDOI}
\PrintBackRefs{\CurrentBib}

\bibitem [\protect \citeauthoryear {%
Katz%
}{%
Katz%
}{%
{\protect \APACyear {2008}}%
}]{%
katz_magma_2008}
\APACinsertmetastar {%
katz_magma_2008}%
\begin{APACrefauthors}%
Katz, R\BPBI F.%
\end{APACrefauthors}%
\unskip\
\newblock
\APACrefYearMonthDay{2008}{{\APACmonth{12}}}{}.
\newblock
{\BBOQ}\APACrefatitle {Magma {Dynamics} with the {Enthalpy} {Method}:
  {Benchmark} {Solutions} and {Magmatic} {Focusing} at {Mid}-ocean {Ridges}}
  {Magma {Dynamics} with the {Enthalpy} {Method}: {Benchmark} {Solutions} and
  {Magmatic} {Focusing} at {Mid}-ocean {Ridges}}.{\BBCQ}
\newblock
\APACjournalVolNumPages{Journal of Petrology}{49}{12}{2099--2121}.
\newblock
\begin{APACrefURL}
  [{2020-01-10}]\url{https://academic.oup.com/petrology/article/49/12/2099/1531301}
  \end{APACrefURL}
\newblock
\begin{APACrefDOI} \doi{10.1093/petrology/egn058} \end{APACrefDOI}
\PrintBackRefs{\CurrentBib}

\bibitem [\protect \citeauthoryear {%
Katz%
}{%
Katz%
}{%
{\protect \APACyear {2010}}%
}]{%
katz_porosity-driven_2010}
\APACinsertmetastar {%
katz_porosity-driven_2010}%
\begin{APACrefauthors}%
Katz, R\BPBI F.%
\end{APACrefauthors}%
\unskip\
\newblock
\APACrefYearMonthDay{2010}{{\APACmonth{11}}}{}.
\newblock
{\BBOQ}\APACrefatitle {Porosity-driven convection and asymmetry beneath
  mid-ocean ridges} {Porosity-driven convection and asymmetry beneath mid-ocean
  ridges}.{\BBCQ}
\newblock
\APACjournalVolNumPages{Geochemistry, Geophysics, Geosystems}{11}{11}{}.
\newblock
\begin{APACrefURL}
  [{2018-11-19}]\url{https://agupubs.onlinelibrary.wiley.com/doi/abs/10.1029/2010GC003282}
  \end{APACrefURL}
\newblock
\begin{APACrefDOI} \doi{10.1029/2010GC003282} \end{APACrefDOI}
\PrintBackRefs{\CurrentBib}

\bibitem [\protect \citeauthoryear {%
Katz%
, Knepley%
, Smith%
, Spiegelman%
\BCBL {}\ \BBA {} Coon%
}{%
Katz%
\ \protect \BOthers {.}}{%
{\protect \APACyear {2007}}%
}]{%
katz_numerical_2007}
\APACinsertmetastar {%
katz_numerical_2007}%
\begin{APACrefauthors}%
Katz, R\BPBI F.%
, Knepley, M\BPBI G.%
, Smith, B.%
, Spiegelman, M.%
\BCBL {}\ \BBA {} Coon, E\BPBI T.%
\end{APACrefauthors}%
\unskip\
\newblock
\APACrefYearMonthDay{2007}{{\APACmonth{08}}}{}.
\newblock
{\BBOQ}\APACrefatitle {Numerical simulation of geodynamic processes with the
  {Portable} {Extensible} {Toolkit} for {Scientific} {Computation}} {Numerical
  simulation of geodynamic processes with the {Portable} {Extensible} {Toolkit}
  for {Scientific} {Computation}}.{\BBCQ}
\newblock
\APACjournalVolNumPages{Physics of the Earth and Planetary
  Interiors}{163}{1}{52--68}.
\newblock
\begin{APACrefURL}
  [{2020-06-24}]\url{http://www.sciencedirect.com/science/article/pii/S0031920107000957}
  \end{APACrefURL}
\newblock
\begin{APACrefDOI} \doi{10.1016/j.pepi.2007.04.016} \end{APACrefDOI}
\PrintBackRefs{\CurrentBib}

\bibitem [\protect \citeauthoryear {%
Katz%
\ \BBA {} Weatherley%
}{%
Katz%
\ \BBA {} Weatherley%
}{%
{\protect \APACyear {2012}}%
}]{%
katz_consequences_2012}
\APACinsertmetastar {%
katz_consequences_2012}%
\begin{APACrefauthors}%
Katz, R\BPBI F.%
\BCBT {}\ \BBA {} Weatherley, S\BPBI M.%
\end{APACrefauthors}%
\unskip\
\newblock
\APACrefYearMonthDay{2012}{{\APACmonth{06}}}{}.
\newblock
{\BBOQ}\APACrefatitle {Consequences of mantle heterogeneity for melt extraction
  at mid-ocean ridges} {Consequences of mantle heterogeneity for melt
  extraction at mid-ocean ridges}.{\BBCQ}
\newblock
\APACjournalVolNumPages{Earth and Planetary Science
  Letters}{335-336}{}{226--237}.
\newblock
\begin{APACrefURL}
  [{2020-05-19}]\url{http://www.sciencedirect.com/science/article/pii/S0012821X12002105}
  \end{APACrefURL}
\newblock
\begin{APACrefDOI} \doi{10.1016/j.epsl.2012.04.042} \end{APACrefDOI}
\PrintBackRefs{\CurrentBib}

\bibitem [\protect \citeauthoryear {%
Kelemen%
, Shimizu%
\BCBL {}\ \BBA {} Salters%
}{%
Kelemen%
\ \protect \BOthers {.}}{%
{\protect \APACyear {1995}}%
}]{%
kelemen_extraction_1995}
\APACinsertmetastar {%
kelemen_extraction_1995}%
\begin{APACrefauthors}%
Kelemen, P\BPBI B.%
, Shimizu, N.%
\BCBL {}\ \BBA {} Salters, V\BPBI J\BPBI M.%
\end{APACrefauthors}%
\unskip\
\newblock
\APACrefYearMonthDay{1995}{{\APACmonth{06}}}{}.
\newblock
{\BBOQ}\APACrefatitle {Extraction of mid-ocean-ridge basalt from the upwelling
  mantle by focused flow of melt in dunite channels} {Extraction of
  mid-ocean-ridge basalt from the upwelling mantle by focused flow of melt in
  dunite channels}.{\BBCQ}
\newblock
\APACjournalVolNumPages{Nature}{375}{6534}{747--753}.
\newblock
\begin{APACrefURL} [{2020-06-01}]\url{https://www.nature.com/articles/375747a0}
  \end{APACrefURL}
\newblock
\APACrefnote{Number: 6534 Publisher: Nature Publishing Group}
\newblock
\begin{APACrefDOI} \doi{10.1038/375747a0} \end{APACrefDOI}
\PrintBackRefs{\CurrentBib}

\bibitem [\protect \citeauthoryear {%
Keszthelyi%
\ \protect \BOthers {.}}{%
Keszthelyi%
\ \protect \BOthers {.}}{%
{\protect \APACyear {2007}}%
}]{%
keszthelyi_new_2007}
\APACinsertmetastar {%
keszthelyi_new_2007}%
\begin{APACrefauthors}%
Keszthelyi, L.%
, Jaeger, W.%
, Milazzo, M.%
, Radebaugh, J.%
, Davies, A\BPBI G.%
\BCBL {}\ \BBA {} Mitchell, K\BPBI L.%
\end{APACrefauthors}%
\unskip\
\newblock
\APACrefYearMonthDay{2007}{{\APACmonth{12}}}{}.
\newblock
{\BBOQ}\APACrefatitle {New estimates for {Io} eruption temperatures:
  {Implications} for the interior} {New estimates for {Io} eruption
  temperatures: {Implications} for the interior}.{\BBCQ}
\newblock
\APACjournalVolNumPages{Icarus}{192}{2}{491--502}.
\newblock
\begin{APACrefURL}
  \url{http://www.sciencedirect.com/science/article/pii/S0019103507003132}
  \end{APACrefURL}
\newblock
\begin{APACrefDOI} \doi{10.1016/j.icarus.2007.07.008} \end{APACrefDOI}
\PrintBackRefs{\CurrentBib}

\bibitem [\protect \citeauthoryear {%
Keszthelyi%
, Jaeger%
, Turtle%
, Milazzo%
\BCBL {}\ \BBA {} Radebaugh%
}{%
Keszthelyi%
\ \protect \BOthers {.}}{%
{\protect \APACyear {2004}}%
}]{%
keszthelyi_post-galileo_2004}
\APACinsertmetastar {%
keszthelyi_post-galileo_2004}%
\begin{APACrefauthors}%
Keszthelyi, L.%
, Jaeger, W\BPBI L.%
, Turtle, E\BPBI P.%
, Milazzo, M.%
\BCBL {}\ \BBA {} Radebaugh, J.%
\end{APACrefauthors}%
\unskip\
\newblock
\APACrefYearMonthDay{2004}{{\APACmonth{05}}}{}.
\newblock
{\BBOQ}\APACrefatitle {A post-{Galileo} view of {Io}'s interior} {A
  post-{Galileo} view of {Io}'s interior}.{\BBCQ}
\newblock
\APACjournalVolNumPages{Icarus}{169}{1}{271--286}.
\newblock
\begin{APACrefURL}
  [{2020-04-21}]\url{http://www.sciencedirect.com/science/article/pii/S0019103504000351}
  \end{APACrefURL}
\newblock
\begin{APACrefDOI} \doi{10.1016/j.icarus.2004.01.005} \end{APACrefDOI}
\PrintBackRefs{\CurrentBib}

\bibitem [\protect \citeauthoryear {%
Keszthelyi%
\ \BBA {} McEwen%
}{%
Keszthelyi%
\ \BBA {} McEwen%
}{%
{\protect \APACyear {1997}}%
}]{%
keszthelyi_magmatic_1997}
\APACinsertmetastar {%
keszthelyi_magmatic_1997}%
\begin{APACrefauthors}%
Keszthelyi, L.%
\BCBT {}\ \BBA {} McEwen, A.%
\end{APACrefauthors}%
\unskip\
\newblock
\APACrefYearMonthDay{1997}{{\APACmonth{12}}}{}.
\newblock
{\BBOQ}\APACrefatitle {Magmatic {Differentiation} of {Io}} {Magmatic
  {Differentiation} of {Io}}.{\BBCQ}
\newblock
\APACjournalVolNumPages{Icarus}{130}{2}{437--448}.
\newblock
\begin{APACrefURL}
  \url{http://www.sciencedirect.com/science/article/pii/S0019103597958371}
  \end{APACrefURL}
\newblock
\begin{APACrefDOI} \doi{10.1006/icar.1997.5837} \end{APACrefDOI}
\PrintBackRefs{\CurrentBib}

\bibitem [\protect \citeauthoryear {%
Keszthelyi%
, McEwen%
\BCBL {}\ \BBA {} Taylor%
}{%
Keszthelyi%
\ \protect \BOthers {.}}{%
{\protect \APACyear {1999}}%
}]{%
keszthelyi_revisiting_1999}
\APACinsertmetastar {%
keszthelyi_revisiting_1999}%
\begin{APACrefauthors}%
Keszthelyi, L.%
, McEwen, A\BPBI S.%
\BCBL {}\ \BBA {} Taylor, G\BPBI J.%
\end{APACrefauthors}%
\unskip\
\newblock
\APACrefYearMonthDay{1999}{{\APACmonth{10}}}{}.
\newblock
{\BBOQ}\APACrefatitle {Revisiting the {Hypothesis} of a {Mushy} {Global}
  {Magma} {Ocean} in {Io}} {Revisiting the {Hypothesis} of a {Mushy} {Global}
  {Magma} {Ocean} in {Io}}.{\BBCQ}
\newblock
\APACjournalVolNumPages{Icarus}{141}{2}{415--419}.
\newblock
\begin{APACrefURL}
  [{2020-05-15}]\url{http://www.sciencedirect.com/science/article/pii/S0019103599961791}
  \end{APACrefURL}
\newblock
\begin{APACrefDOI} \doi{10.1006/icar.1999.6179} \end{APACrefDOI}
\PrintBackRefs{\CurrentBib}

\bibitem [\protect \citeauthoryear {%
Khurana%
\ \protect \BOthers {.}}{%
Khurana%
\ \protect \BOthers {.}}{%
{\protect \APACyear {2011}}%
}]{%
khurana_evidence_2011}
\APACinsertmetastar {%
khurana_evidence_2011}%
\begin{APACrefauthors}%
Khurana, K\BPBI K.%
, Jia, X.%
, Kivelson, M\BPBI G.%
, Nimmo, F.%
, Schubert, G.%
\BCBL {}\ \BBA {} Russell, C\BPBI T.%
\end{APACrefauthors}%
\unskip\
\newblock
\APACrefYearMonthDay{2011}{{\APACmonth{06}}}{}.
\newblock
{\BBOQ}\APACrefatitle {Evidence of a {Global} {Magma} {Ocean} in {Io}'s
  {Interior}} {Evidence of a {Global} {Magma} {Ocean} in {Io}'s
  {Interior}}.{\BBCQ}
\newblock
\APACjournalVolNumPages{Science}{332}{6034}{1186--1189}.
\newblock
\begin{APACrefURL}
  [{2017-10-11}]\url{http://science.sciencemag.org/content/332/6034/1186}
  \end{APACrefURL}
\newblock
\begin{APACrefDOI} \doi{10.1126/science.1201425} \end{APACrefDOI}
\PrintBackRefs{\CurrentBib}

\bibitem [\protect \citeauthoryear {%
Kirchoff%
, McKinnon%
\BCBL {}\ \BBA {} Schenk%
}{%
Kirchoff%
\ \protect \BOthers {.}}{%
{\protect \APACyear {2011}}%
}]{%
kirchoff_global_2011}
\APACinsertmetastar {%
kirchoff_global_2011}%
\begin{APACrefauthors}%
Kirchoff, M\BPBI R.%
, McKinnon, W\BPBI B.%
\BCBL {}\ \BBA {} Schenk, P\BPBI M.%
\end{APACrefauthors}%
\unskip\
\newblock
\APACrefYearMonthDay{2011}{{\APACmonth{01}}}{}.
\newblock
{\BBOQ}\APACrefatitle {Global distribution of volcanic centers and mountains on
  {Io}: {Control} by asthenospheric heating and implications for mountain
  formation} {Global distribution of volcanic centers and mountains on {Io}:
  {Control} by asthenospheric heating and implications for mountain
  formation}.{\BBCQ}
\newblock
\APACjournalVolNumPages{Earth and Planetary Science Letters}{301}{1}{22--30}.
\newblock
\begin{APACrefURL}
  \url{http://www.sciencedirect.com/science/article/pii/S0012821X10007132}
  \end{APACrefURL}
\newblock
\begin{APACrefDOI} \doi{10.1016/j.epsl.2010.11.018} \end{APACrefDOI}
\PrintBackRefs{\CurrentBib}

\bibitem [\protect \citeauthoryear {%
Lainey%
, Arlot%
, Karatekin%
\BCBL {}\ \BBA {} Van~Hoolst%
}{%
Lainey%
\ \protect \BOthers {.}}{%
{\protect \APACyear {2009}}%
}]{%
lainey_strong_2009}
\APACinsertmetastar {%
lainey_strong_2009}%
\begin{APACrefauthors}%
Lainey, V.%
, Arlot, J\BHBI E.%
, Karatekin, {\textbackslash}.%
\BCBL {}\ \BBA {} Van~Hoolst, T.%
\end{APACrefauthors}%
\unskip\
\newblock
\APACrefYearMonthDay{2009}{{\APACmonth{06}}}{}.
\newblock
{\BBOQ}\APACrefatitle {Strong tidal dissipation in {Io} and {Jupiter} from
  astrometric observations} {Strong tidal dissipation in {Io} and {Jupiter}
  from astrometric observations}.{\BBCQ}
\newblock
\APACjournalVolNumPages{Nature}{459}{7249}{957--959}.
\newblock
\begin{APACrefURL}
  [{2017-08-14}]\url{http://www.nature.com/nature/journal/v459/n7249/full/nature08108.html?foxtrotcallback=true}
  \end{APACrefURL}
\newblock
\begin{APACrefDOI} \doi{10.1038/nature08108} \end{APACrefDOI}
\PrintBackRefs{\CurrentBib}

\bibitem [\protect \citeauthoryear {%
McEwen%
\ \protect \BOthers {.}}{%
McEwen%
\ \protect \BOthers {.}}{%
{\protect \APACyear {1998}}%
}]{%
mcewen_high-temperature_1998}
\APACinsertmetastar {%
mcewen_high-temperature_1998}%
\begin{APACrefauthors}%
McEwen, A\BPBI S.%
, Keszthelyi, L.%
, Spencer, J\BPBI R.%
, Schubert, G.%
, Matson, D\BPBI L.%
, Lopes-Gautier, R.%
\BDBL {}Belton, M\BPBI J\BPBI S.%
\end{APACrefauthors}%
\unskip\
\newblock
\APACrefYearMonthDay{1998}{{\APACmonth{07}}}{}.
\newblock
{\BBOQ}\APACrefatitle {High-{Temperature} {Silicate} {Volcanism} on {Jupiter}'s
  {Moon} {Io}} {High-{Temperature} {Silicate} {Volcanism} on {Jupiter}'s {Moon}
  {Io}}.{\BBCQ}
\newblock
\APACjournalVolNumPages{Science}{281}{5373}{87--90}.
\newblock
\begin{APACrefURL}
  [{2017-10-20}]\url{http://science.sciencemag.org/content/281/5373/87}
  \end{APACrefURL}
\newblock
\begin{APACrefDOI} \doi{10.1126/science.281.5373.87} \end{APACrefDOI}
\PrintBackRefs{\CurrentBib}

\bibitem [\protect \citeauthoryear {%
McKenzie%
}{%
McKenzie%
}{%
{\protect \APACyear {1984}}%
}]{%
mckenzie_generation_1984}
\APACinsertmetastar {%
mckenzie_generation_1984}%
\begin{APACrefauthors}%
McKenzie, D.%
\end{APACrefauthors}%
\unskip\
\newblock
\APACrefYearMonthDay{1984}{{\APACmonth{08}}}{}.
\newblock
{\BBOQ}\APACrefatitle {The {Generation} and {Compaction} of {Partially}
  {Molten} {Rock}} {The {Generation} and {Compaction} of {Partially} {Molten}
  {Rock}}.{\BBCQ}
\newblock
\APACjournalVolNumPages{Journal of Petrology}{25}{3}{713--765}.
\newblock
\begin{APACrefURL}
  [{2018-11-01}]\url{https://academic.oup.com/petrology/article/25/3/713/1394279}
  \end{APACrefURL}
\newblock
\begin{APACrefDOI} \doi{10.1093/petrology/25.3.713} \end{APACrefDOI}
\PrintBackRefs{\CurrentBib}

\bibitem [\protect \citeauthoryear {%
Moore%
}{%
Moore%
}{%
{\protect \APACyear {2001}}%
}]{%
moore_thermal_2001}
\APACinsertmetastar {%
moore_thermal_2001}%
\begin{APACrefauthors}%
Moore, W\BPBI B.%
\end{APACrefauthors}%
\unskip\
\newblock
\APACrefYearMonthDay{2001}{{\APACmonth{12}}}{}.
\newblock
{\BBOQ}\APACrefatitle {The {Thermal} {State} of {Io}} {The {Thermal} {State} of
  {Io}}.{\BBCQ}
\newblock
\APACjournalVolNumPages{Icarus}{154}{2}{548--550}.
\newblock
\begin{APACrefURL}
  \url{http://www.sciencedirect.com/science/article/pii/S0019103501967399}
  \end{APACrefURL}
\newblock
\begin{APACrefDOI} \doi{10.1006/icar.2001.6739} \end{APACrefDOI}
\PrintBackRefs{\CurrentBib}

\bibitem [\protect \citeauthoryear {%
Moore%
}{%
Moore%
}{%
{\protect \APACyear {2003}}%
}]{%
moore_tidal_2003}
\APACinsertmetastar {%
moore_tidal_2003}%
\begin{APACrefauthors}%
Moore, W\BPBI B.%
\end{APACrefauthors}%
\unskip\
\newblock
\APACrefYearMonthDay{2003}{{\APACmonth{08}}}{}.
\newblock
{\BBOQ}\APACrefatitle {Tidal heating and convection in {Io}} {Tidal heating and
  convection in {Io}}.{\BBCQ}
\newblock
\APACjournalVolNumPages{Journal of Geophysical Research:
  Planets}{108}{E8}{5096}.
\newblock
\begin{APACrefURL}
  \url{http://onlinelibrary.wiley.com/doi/10.1029/2002JE001943/abstract}
  \end{APACrefURL}
\newblock
\begin{APACrefDOI} \doi{10.1029/2002JE001943} \end{APACrefDOI}
\PrintBackRefs{\CurrentBib}

\bibitem [\protect \citeauthoryear {%
O'Reilly%
\ \BBA {} Davies%
}{%
O'Reilly%
\ \BBA {} Davies%
}{%
{\protect \APACyear {1981}}%
}]{%
oreilly_magma_1981}
\APACinsertmetastar {%
oreilly_magma_1981}%
\begin{APACrefauthors}%
O'Reilly, T\BPBI C.%
\BCBT {}\ \BBA {} Davies, G\BPBI F.%
\end{APACrefauthors}%
\unskip\
\newblock
\APACrefYearMonthDay{1981}{{\APACmonth{04}}}{}.
\newblock
{\BBOQ}\APACrefatitle {Magma transport of heat on {Io}: {A} mechanism allowing
  a thick lithosphere} {Magma transport of heat on {Io}: {A} mechanism allowing
  a thick lithosphere}.{\BBCQ}
\newblock
\APACjournalVolNumPages{Geophysical Research Letters}{8}{4}{313--316}.
\newblock
\begin{APACrefURL}
  \url{http://onlinelibrary.wiley.com/doi/10.1029/GL008i004p00313/abstract}
  \end{APACrefURL}
\newblock
\begin{APACrefDOI} \doi{10.1029/GL008i004p00313} \end{APACrefDOI}
\PrintBackRefs{\CurrentBib}

\bibitem [\protect \citeauthoryear {%
Peale%
, Cassen%
\BCBL {}\ \BBA {} Reynolds%
}{%
Peale%
\ \protect \BOthers {.}}{%
{\protect \APACyear {1979}}%
}]{%
peale_melting_1979}
\APACinsertmetastar {%
peale_melting_1979}%
\begin{APACrefauthors}%
Peale, S\BPBI J.%
, Cassen, P.%
\BCBL {}\ \BBA {} Reynolds, R\BPBI T.%
\end{APACrefauthors}%
\unskip\
\newblock
\APACrefYearMonthDay{1979}{{\APACmonth{03}}}{}.
\newblock
{\BBOQ}\APACrefatitle {Melting of {Io} by {Tidal} {Dissipation}} {Melting of
  {Io} by {Tidal} {Dissipation}}.{\BBCQ}
\newblock
\APACjournalVolNumPages{Science}{203}{4383}{892--894}.
\newblock
\begin{APACrefURL}
  [{2020-04-23}]\url{https://science.sciencemag.org/content/203/4383/892}
  \end{APACrefURL}
\newblock
\APACrefnote{Publisher: American Association for the Advancement of Science
  Section: Reports}
\newblock
\begin{APACrefDOI} \doi{10.1126/science.203.4383.892} \end{APACrefDOI}
\PrintBackRefs{\CurrentBib}

\bibitem [\protect \citeauthoryear {%
Rees~Jones%
\ \BBA {} Katz%
}{%
Rees~Jones%
\ \BBA {} Katz%
}{%
{\protect \APACyear {2018}}%
}]{%
rees_jones_reaction-infiltration_2018}
\APACinsertmetastar {%
rees_jones_reaction-infiltration_2018}%
\begin{APACrefauthors}%
Rees~Jones, D\BPBI W.%
\BCBT {}\ \BBA {} Katz, R\BPBI F.%
\end{APACrefauthors}%
\unskip\
\newblock
\APACrefYearMonthDay{2018}{{\APACmonth{10}}}{}.
\newblock
{\BBOQ}\APACrefatitle {Reaction-infiltration instability in a compacting porous
  medium} {Reaction-infiltration instability in a compacting porous
  medium}.{\BBCQ}
\newblock
\APACjournalVolNumPages{Journal of Fluid Mechanics}{852}{}{5--36}.
\newblock
\begin{APACrefURL}
  [{2020-05-20}]\url{https://www.cambridge.org/core/journals/journal-of-fluid-mechanics/article/reactioninfiltration-instability-in-a-compacting-porous-medium/ED989F0A57DDEFED45573509DCCA4337}
  \end{APACrefURL}
\newblock
\APACrefnote{Publisher: Cambridge University Press}
\newblock
\begin{APACrefDOI} \doi{10.1017/jfm.2018.524} \end{APACrefDOI}
\PrintBackRefs{\CurrentBib}

\bibitem [\protect \citeauthoryear {%
Renaud%
\ \BBA {} Henning%
}{%
Renaud%
\ \BBA {} Henning%
}{%
{\protect \APACyear {2018}}%
}]{%
renaud_increased_2018}
\APACinsertmetastar {%
renaud_increased_2018}%
\begin{APACrefauthors}%
Renaud, J\BPBI P.%
\BCBT {}\ \BBA {} Henning, W\BPBI G.%
\end{APACrefauthors}%
\unskip\
\newblock
\APACrefYearMonthDay{2018}{{\APACmonth{04}}}{}.
\newblock
{\BBOQ}\APACrefatitle {Increased {Tidal} {Dissipation} {Using} {Advanced}
  {Rheological} {Models}: {Implications} for {Io} and {Tidally} {Active}
  {Exoplanets}} {Increased {Tidal} {Dissipation} {Using} {Advanced}
  {Rheological} {Models}: {Implications} for {Io} and {Tidally} {Active}
  {Exoplanets}}.{\BBCQ}
\newblock
\APACjournalVolNumPages{The Astrophysical Journal}{857}{2}{98}.
\newblock
\begin{APACrefURL}
  [{2019-04-26}]\url{https://doi.org/10.3847%2F1538-4357%2Faab784}
  \end{APACrefURL}
\newblock
\begin{APACrefDOI} \doi{10.3847/1538-4357/aab784} \end{APACrefDOI}
\PrintBackRefs{\CurrentBib}

\bibitem [\protect \citeauthoryear {%
Ross%
, Schubert%
, Spohn%
\BCBL {}\ \BBA {} Gaskell%
}{%
Ross%
\ \protect \BOthers {.}}{%
{\protect \APACyear {1990}}%
}]{%
ross_internal_1990}
\APACinsertmetastar {%
ross_internal_1990}%
\begin{APACrefauthors}%
Ross, M\BPBI N.%
, Schubert, G.%
, Spohn, T.%
\BCBL {}\ \BBA {} Gaskell, R\BPBI W.%
\end{APACrefauthors}%
\unskip\
\newblock
\APACrefYearMonthDay{1990}{{\APACmonth{06}}}{}.
\newblock
{\BBOQ}\APACrefatitle {Internal structure of {Io} and the global distribution
  of its topography} {Internal structure of {Io} and the global distribution of
  its topography}.{\BBCQ}
\newblock
\APACjournalVolNumPages{Icarus}{85}{2}{309--325}.
\newblock
\begin{APACrefURL}
  \url{http://www.sciencedirect.com/science/article/pii/001910359090119T}
  \end{APACrefURL}
\newblock
\begin{APACrefDOI} \doi{≈} \end{APACrefDOI}
\PrintBackRefs{\CurrentBib}

\bibitem [\protect \citeauthoryear {%
Roth%
\ \protect \BOthers {.}}{%
Roth%
\ \protect \BOthers {.}}{%
{\protect \APACyear {2017}}%
}]{%
roth_constraints_2017}
\APACinsertmetastar {%
roth_constraints_2017}%
\begin{APACrefauthors}%
Roth, L.%
, Saur, J.%
, Retherford, K\BPBI D.%
, Blöcker, A.%
, Strobel, D\BPBI F.%
\BCBL {}\ \BBA {} Feldman, P\BPBI D.%
\end{APACrefauthors}%
\unskip\
\newblock
\APACrefYearMonthDay{2017}{}{}.
\newblock
{\BBOQ}\APACrefatitle {Constraints on {Io}'s interior from auroral spot
  oscillations} {Constraints on {Io}'s interior from auroral spot
  oscillations}.{\BBCQ}
\newblock
\APACjournalVolNumPages{Journal of Geophysical Research: Space
  Physics}{122}{2}{1903--1927}.
\newblock
\begin{APACrefURL}
  [{2020-06-01}]\url{https://agupubs.onlinelibrary.wiley.com/doi/abs/10.1002/2016JA023701}
  \end{APACrefURL}
\newblock
\APACrefnote{\_eprint:
  https://agupubs.onlinelibrary.wiley.com/doi/pdf/10.1002/2016JA023701}
\newblock
\begin{APACrefDOI} \doi{10.1002/2016JA023701} \end{APACrefDOI}
\PrintBackRefs{\CurrentBib}

\bibitem [\protect \citeauthoryear {%
Segatz%
, Spohn%
, Ross%
\BCBL {}\ \BBA {} Schubert%
}{%
Segatz%
\ \protect \BOthers {.}}{%
{\protect \APACyear {1988}}%
}]{%
segatz_tidal_1988}
\APACinsertmetastar {%
segatz_tidal_1988}%
\begin{APACrefauthors}%
Segatz, M.%
, Spohn, T.%
, Ross, M\BPBI N.%
\BCBL {}\ \BBA {} Schubert, G.%
\end{APACrefauthors}%
\unskip\
\newblock
\APACrefYearMonthDay{1988}{{\APACmonth{08}}}{}.
\newblock
{\BBOQ}\APACrefatitle {Tidal dissipation, surface heat flow, and figure of
  viscoelastic models of {Io}} {Tidal dissipation, surface heat flow, and
  figure of viscoelastic models of {Io}}.{\BBCQ}
\newblock
\APACjournalVolNumPages{Icarus}{75}{2}{187--206}.
\newblock
\begin{APACrefURL}
  \url{http://www.sciencedirect.com/science/article/pii/0019103588900012}
  \end{APACrefURL}
\newblock
\begin{APACrefDOI} \doi{10.1016/0019-1035(88)90001-2} \end{APACrefDOI}
\PrintBackRefs{\CurrentBib}

\bibitem [\protect \citeauthoryear {%
Spencer%
, Katz%
\BCBL {}\ \BBA {} Hewitt%
}{%
Spencer%
, Katz%
\BCBL {}\ \BBA {} Hewitt%
}{%
{\protect \APACyear {2020}}%
}]{%
spencer_magmatic_2020}
\APACinsertmetastar {%
spencer_magmatic_2020}%
\begin{APACrefauthors}%
Spencer, D\BPBI C.%
, Katz, R\BPBI F.%
\BCBL {}\ \BBA {} Hewitt, I\BPBI J.%
\end{APACrefauthors}%
\unskip\
\newblock
\APACrefYearMonthDay{2020}{}{}.
\newblock
{\BBOQ}\APACrefatitle {Magmatic {Intrusions} {Control} {Io}'s {Crustal}
  {Thickness}} {Magmatic {Intrusions} {Control} {Io}'s {Crustal}
  {Thickness}}.{\BBCQ}
\newblock
\APACjournalVolNumPages{Journal of Geophysical Research:
  Planets}{125}{6}{e2020JE006443}.
\newblock
\begin{APACrefURL}
  [{2020-06-15}]\url{https://agupubs.onlinelibrary.wiley.com/doi/abs/10.1029/2020JE006443}
  \end{APACrefURL}
\newblock
\APACrefnote{\_eprint:
  https://agupubs.onlinelibrary.wiley.com/doi/pdf/10.1029/2020JE006443}
\newblock
\begin{APACrefDOI} \doi{10.1029/2020JE006443} \end{APACrefDOI}
\PrintBackRefs{\CurrentBib}

\bibitem [\protect \citeauthoryear {%
Spencer%
, Katz%
, Hewitt%
\BCBL {}\ \BBA {} May%
}{%
Spencer%
, Katz%
, Hewitt%
\BCBL {}\ \BBA {} May%
}{%
{\protect \APACyear {2020}}%
}]{%
spencer_spencer-spaceio_composition_2020}
\APACinsertmetastar {%
spencer_spencer-spaceio_composition_2020}%
\begin{APACrefauthors}%
Spencer, D\BPBI C.%
, Katz, R\BPBI F.%
, Hewitt, I\BPBI J.%
\BCBL {}\ \BBA {} May, D.%
\end{APACrefauthors}%
\unskip\
\newblock
\APACrefYearMonthDay{2020}{{\APACmonth{06}}}{}.
\newblock
\APACrefbtitle {Spencer-space/io\_composition: {First} release of {IoComp}.}
  {Spencer-space/io\_composition: {First} release of {IoComp}.}
\newblock
\APACaddressPublisher{}{Zenodo}.
\newblock
\begin{APACrefURL} [{2020-06-17}]\url{https://zenodo.org/record/3898245}
  \end{APACrefURL}
\newblock
\begin{APACrefDOI} \doi{10.5281/zenodo.3898245} \end{APACrefDOI}
\PrintBackRefs{\CurrentBib}

\bibitem [\protect \citeauthoryear {%
Steinke%
, Hu%
, Höning%
, van~der Wal%
\BCBL {}\ \BBA {} Vermeersen%
}{%
Steinke%
\ \protect \BOthers {.}}{%
{\protect \APACyear {2020}}%
}]{%
steinke_tidally_2020}
\APACinsertmetastar {%
steinke_tidally_2020}%
\begin{APACrefauthors}%
Steinke, T.%
, Hu, H.%
, Höning, D.%
, van~der Wal, W.%
\BCBL {}\ \BBA {} Vermeersen, B.%
\end{APACrefauthors}%
\unskip\
\newblock
\APACrefYearMonthDay{2020}{{\APACmonth{01}}}{}.
\newblock
{\BBOQ}\APACrefatitle {Tidally induced lateral variations of {Io}'s interior}
  {Tidally induced lateral variations of {Io}'s interior}.{\BBCQ}
\newblock
\APACjournalVolNumPages{Icarus}{335}{}{113299}.
\newblock
\begin{APACrefURL}
  [{2020-01-10}]\url{http://www.sciencedirect.com/science/article/pii/S0019103518307632}
  \end{APACrefURL}
\newblock
\begin{APACrefDOI} \doi{10.1016/j.icarus.2019.05.001} \end{APACrefDOI}
\PrintBackRefs{\CurrentBib}

\bibitem [\protect \citeauthoryear {%
Tobie%
, Mocquet%
\BCBL {}\ \BBA {} Sotin%
}{%
Tobie%
\ \protect \BOthers {.}}{%
{\protect \APACyear {2005}}%
}]{%
tobie_tidal_2005}
\APACinsertmetastar {%
tobie_tidal_2005}%
\begin{APACrefauthors}%
Tobie, G.%
, Mocquet, A.%
\BCBL {}\ \BBA {} Sotin, C.%
\end{APACrefauthors}%
\unskip\
\newblock
\APACrefYearMonthDay{2005}{{\APACmonth{10}}}{}.
\newblock
{\BBOQ}\APACrefatitle {Tidal dissipation within large icy satellites:
  {Applications} to {Europa} and {Titan}} {Tidal dissipation within large icy
  satellites: {Applications} to {Europa} and {Titan}}.{\BBCQ}
\newblock
\APACjournalVolNumPages{Icarus}{177}{2}{534--549}.
\newblock
\begin{APACrefURL}
  \url{http://www.sciencedirect.com/science/article/pii/S0019103505001582}
  \end{APACrefURL}
\newblock
\begin{APACrefDOI} \doi{10.1016/j.icarus.2005.04.006} \end{APACrefDOI}
\PrintBackRefs{\CurrentBib}

\bibitem [\protect \citeauthoryear {%
Turcotte%
\ \BBA {} Schubert%
}{%
Turcotte%
\ \BBA {} Schubert%
}{%
{\protect \APACyear {2014}}%
}]{%
turcotte_geodynamics_2014}
\APACinsertmetastar {%
turcotte_geodynamics_2014}%
\begin{APACrefauthors}%
Turcotte, D.%
\BCBT {}\ \BBA {} Schubert, G.%
\end{APACrefauthors}%
\unskip\
\newblock
\APACrefYear{2014}.
\newblock
\APACrefbtitle {Geodynamics} {Geodynamics}\ (\PrintOrdinal{3}\ \BEd).
\newblock
\APACaddressPublisher{Cambridge}{Cambridge University Press}.
\newblock
\begin{APACrefURL}
  [{2020-07-22}]\url{https://www.cambridge.org/core/books/geodynamics/E0E847DA9FE68BDB90C2E457791F0C98}
  \end{APACrefURL}
\newblock
\begin{APACrefDOI} \doi{10.1017/CBO9780511843877} \end{APACrefDOI}
\PrintBackRefs{\CurrentBib}

\bibitem [\protect \citeauthoryear {%
Tyler%
, Henning%
\BCBL {}\ \BBA {} Hamilton%
}{%
Tyler%
\ \protect \BOthers {.}}{%
{\protect \APACyear {2015}}%
}]{%
tyler_tidal_2015}
\APACinsertmetastar {%
tyler_tidal_2015}%
\begin{APACrefauthors}%
Tyler, R\BPBI H.%
, Henning, W\BPBI G.%
\BCBL {}\ \BBA {} Hamilton, C\BPBI W.%
\end{APACrefauthors}%
\unskip\
\newblock
\APACrefYearMonthDay{2015}{}{}.
\newblock
{\BBOQ}\APACrefatitle {Tidal {Heating} in a {Magma} {Ocean} within {Jupiter}'s
  {Moon} {Io}} {Tidal {Heating} in a {Magma} {Ocean} within {Jupiter}'s {Moon}
  {Io}}.{\BBCQ}
\newblock
\APACjournalVolNumPages{The Astrophysical Journal Supplement
  Series}{218}{2}{22}.
\newblock
\begin{APACrefURL} \url{http://stacks.iop.org/0067-0049/218/i=2/a=22}
  \end{APACrefURL}
\newblock
\begin{APACrefDOI} \doi{10.1088/0067-0049/218/2/22} \end{APACrefDOI}
\PrintBackRefs{\CurrentBib}

\bibitem [\protect \citeauthoryear {%
Williams%
\ \protect \BOthers {.}}{%
Williams%
\ \protect \BOthers {.}}{%
{\protect \APACyear {2011}}%
}]{%
williams_volcanism_2011}
\APACinsertmetastar {%
williams_volcanism_2011}%
\begin{APACrefauthors}%
Williams, D\BPBI A.%
, Keszthelyi, L\BPBI P.%
, Crown, D\BPBI A.%
, Yff, J\BPBI A.%
, Jaeger, W\BPBI L.%
, Schenk, P\BPBI M.%
\BDBL {}Becker, T\BPBI L.%
\end{APACrefauthors}%
\unskip\
\newblock
\APACrefYearMonthDay{2011}{{\APACmonth{07}}}{}.
\newblock
{\BBOQ}\APACrefatitle {Volcanism on {Io}: {New} insights from global geologic
  mapping} {Volcanism on {Io}: {New} insights from global geologic
  mapping}.{\BBCQ}
\newblock
\APACjournalVolNumPages{Icarus}{214}{1}{91--112}.
\newblock
\begin{APACrefURL}
  [{2017-10-20}]\url{https://ezproxy-prd.bodleian.ox.ac.uk:2073/science/article/pii/S0019103511001710}
  \end{APACrefURL}
\newblock
\begin{APACrefDOI} \doi{10.1016/j.icarus.2011.05.007} \end{APACrefDOI}
\PrintBackRefs{\CurrentBib}

\bibitem [\protect \citeauthoryear {%
Williams%
, Wilson%
\BCBL {}\ \BBA {} Greeley%
}{%
Williams%
\ \protect \BOthers {.}}{%
{\protect \APACyear {2000}}%
}]{%
williams_komatiite_2000}
\APACinsertmetastar {%
williams_komatiite_2000}%
\begin{APACrefauthors}%
Williams, D\BPBI A.%
, Wilson, A\BPBI H.%
\BCBL {}\ \BBA {} Greeley, R.%
\end{APACrefauthors}%
\unskip\
\newblock
\APACrefYearMonthDay{2000}{}{}.
\newblock
{\BBOQ}\APACrefatitle {A komatiite analog to potential ultramafic materials on
  {Io}} {A komatiite analog to potential ultramafic materials on {Io}}.{\BBCQ}
\newblock
\APACjournalVolNumPages{Journal of Geophysical Research:
  Planets}{105}{E1}{1671--1684}.
\newblock
\begin{APACrefURL}
  [{2020-05-15}]\url{https://agupubs.onlinelibrary.wiley.com/doi/abs/10.1029/1999JE001157}
  \end{APACrefURL}
\newblock
\APACrefnote{\_eprint:
  https://agupubs.onlinelibrary.wiley.com/doi/pdf/10.1029/1999JE001157}
\newblock
\begin{APACrefDOI} \doi{10.1029/1999JE001157} \end{APACrefDOI}
\PrintBackRefs{\CurrentBib}

\end{thebibliography}

%
%
%
%
%

\end{document}